\documentclass[showpacs,preprintnumbers,amsmath,amssymb]{revtex4}

\usepackage{epsfig}
\usepackage{graphicx}
\usepackage{dcolumn}
\usepackage{bm}

\begin{document}


\title{Complex Electronic Behavior in a 1D Correlated Quantum Liquid}
\author{J. M. P. Carmelo}
\affiliation{Department of Physics, Massachusetts Institute of Technology, Cambridge,
Massachusetts 02139-4307, USA}
\date{5 August 2005}


\begin{abstract}
The relation of the rotated-electron site distribution configurations that describe the
energy eigenstates of the one-dimensional Hubbard model to the momentum occupancy
configurations of the same states associated with the Bethe-ansatz quantum numbers is
clarified. This allows an explicit study of the complex behavior that occurs for the
electronic degrees of freedom, following the electron - rotated-electron unitary
transformation. Such a behavior plays a central role in the non-perturbative mechanisms
associated with the organization of the electronic degrees of freedom in terms of exotic
objects whose momentum occupancy configurations describe the energy eigenstates. Our
study involves the introduction of the concepts of {\it local pseudoparticle} and {\it
effective pseudoparticle lattice}, needed for the relation of both the local
pseudoparticle internal structure and the pseudoparticle occupancy configurations to the
rotated-electron site distribution configurations. Our results provide further useful
information about the microscopic mechanisms behind the anomalous finite-energy spectral
properties of low-dimensional complex materials.
\end{abstract}

\pacs{03, 70}

\maketitle
\section{INTRODUCTION}

The one-dimensional (1D) Hubbard model describes successfully the role of electronic
correlations in the unusual electronic and optical properties observed in quasi-1D
Mott-Hubbard insulators \cite{Tohyama}, doped Mott-Hubbard insulators
\cite{Vescoli,super}, and metals \cite{spectral0}. It is one of the few realistic models
for which one can exactly calculate all the energy eigenstates and their energies
\cite{Lieb,Takahashi}.

Recently, the quantum numbers provided by the model Bethe-ansatz solution
\cite{Lieb,Takahashi} and $\eta$-spin and spin $SU(2)$ symmetries \cite{HL,Yang89} were
shown to correspond to occupancy configurations of exotic quantum objects \cite{I}.
However, the relation of the local rotated-electron site distribution configurations that
describe the energy eigenstates of the model to the momentum occupancy configurations of
such objects was not fully clarified. The main goal of this paper is to present a
detailed study of the relation between the rotated-electron site distribution
configurations that emerge from the electron - rotated-electron unitary transformation
\cite{I} and the momentum occupancy configurations of such quantum objects. Clarification
of this problem provides important information about the microscopic mechanisms that
control the unusual finite-energy spectral properties of the model \cite{Penc96-97,V}.
The solution of the problem reported here involves the introduction of the concepts of a
{\it local pseudoparticle} and an {\it effective pseudoparticle lattice}. It is found
that the bare-momentum pseudoparticle description associated with the Bethe-ansatz
Takahasi's thermodynamic equations \cite{Takahashi,I} is related by Fourier transforms to
such a local-pseudoparticle representation. The latter representation refers to
local-pseudoparticle occupancy configurations of spatial coordinates that correspond to
an effective pseudoparticle lattice. Such configurations are expressed here in terms of
the corresponding rotated-electron site distribution configurations. The research
presented in this paper follows the introduction of the complete holon, spinon, and $c$
pseudoparticle description of Ref. \cite{I} and is complementary to the studies on the
finite-energy spectral-weight distributions of Ref. \cite{V}, which are used in Refs.
\cite{super,spectral0,spectral} in the description of the unusual properties observed in
quasi-1D compounds. Indeed, the results obtained here provide important and useful
information for the identification of the scatterer and scattering centers that control
the spectral properties of the model \cite{S}. Furthermore, such results are also of
interest for the understanding of the spectral properties of the new quantum systems
described by cold fermionic atoms in optical lattices with on-site repulsion
\cite{Zoller}.

The results of this paper also provide further useful information about the complex
behavior that occurs for the rotated electrons that emerge from the electron -
rotated-electron unitary transformation. Such a behavior was identified in Ref. \cite{I}
and refers to an exotic spin-charge separation: For the whole Hilbert space and energies,
the spin and charge degrees of freedom of the spin-projection $\sigma$-rotated electrons
of the singly occupied sites separate, giving rise to chargeless spin-projection $\sigma$
spinons and $\eta$-spinless and spinless $c$ pseudoparticles of electronic charge $-e$,
respectively. This is an example of the behavior observed in many complex physical and
biological systems: their parts interact, leading to new properties, absent in the
individual parts.

The paper is organized as follows: In Sec. II we introduce the 1D Hubbard model and
summarize the concept of rotated electron as well as the description of the model in
terms of holons, spinons, and pseudoparticles. In Sec. III we discuss the issue of the
choice of the suitable complete set of energy eigenstates of the model in the limit
$U/t\rightarrow\infty$, where $U$ is the on-site repulsion and $t$ the first-neighbor
transfer integral. This includes the introduction of a set of basic properties which are
useful for finding the internal structure of the local composite $\alpha\nu$
pseudoparticle, which is one of the subjects of Sec. IV. In that section we introduce a
complete basis of local states which we express in terms of rotated-electron local
charge, spin, and $c$ pseudoparticle sequences. Moreover, we find the rotated-electron
site distribution configurations which describe the internal structure of the local
$\alpha\nu$ pseudoparticle. The concepts of a local $c$ pseudoparticle and a $\alpha\nu$
effective pseudoparticle lattice are introduced in Sec. V, where we also express the
energy eigenstates in terms of the Fourier-transform superpositions of local charge,
spin, and $c$ pseudoparticle sequences introduced in the previous section. Finally, in
Sec. VI we present the discussion and concluding remarks.

\section{THE 1D HUBBARD MODEL, ROTATED ELECTRONS, AND SUMMARY OF THE
PSEUDOPARTICLE, HOLON, AND SPINON DESCRIPTION}

In a chemical potential $\mu $ and magnetic field $H$ the 1D Hubbard Hamiltonian can be
written as,

\begin{equation}
\hat{H} = {\hat{H}}_{SO(4)} + \sum_{\alpha=c,s}\mu_{\alpha}\,{\hat{S}}^z_{\alpha}
\label{H} \, ; \hspace{0.5cm} {\hat{H}}_{SO(4)} = {\hat{H}}_H - (U/2)[\hat{N} + N_a/2] \,
; \hspace{0.5cm} {\hat{H}}_H = \hat{T} + U\,\hat{D} \, ,
\end{equation}
where ${\hat{H}}_{SO(4)}$ has $SO(4)$ symmetry \cite{HL,Yang89,I} and ${\hat{H}}_H$ is
the simple Hubbard model. The expressions of the kinetic-energy operator $\hat{T}$ and
electron double-occupation operator $\hat{D}$ are given in Eqs. (5) and (6),
respectively, of Ref. \cite{I}. Moreover, ${\hat{N}}$ and ${\hat{N}}_{\sigma}$ are the
electronic number operators and
$\hat{n}_{j,\,\sigma}=c_{j,\,\sigma}^{\dagger}\,c_{j,\,\sigma}$. The number of lattice
sites $N_a$ is even and large and we consider periodic boundary conditions. We denote by
$N_{\uparrow}$ and $N_{\downarrow}$ the number of spin-up electrons and spin-down
electrons, respectively, and by $N=N_{\uparrow}+N_{\downarrow}$ the number of electrons.
The lattice constant is denoted by $a$ and thus the length of the system is $L=N_a\,a$.
We consider electronic densities and spin densities given by $n=n_{\uparrow
}+n_{\downarrow}$ and $m=n_{\uparrow}-n_{\downarrow}$, respectively, where
$n_{\sigma}=N_{\sigma}/L$ and $n=N/L$. These densities belong to the domains defined by
the following inequalities $0\leq na \leq 1$\, ; $1\leq na \leq 2$ and $-na\leq ma \leq
na$\, ; $-(2-na)\leq ma \leq (2-na)$, respectively. The momentum operator reads,

\begin{equation}
\hat{P} = \sum_{\sigma=\uparrow ,\,\downarrow }\sum_{k}\, \hat{N}_{\sigma} (k)\, k =
{L\over 2\pi}\sum_{\sigma=\uparrow ,\,\downarrow }\,\int_{-\pi/a}^{+\pi/a} dk\,
\hat{N}_{\sigma} (k)\, k \, , \label{Popel}
\end{equation}
and commutes with the Hamiltonians introduced in Eq. (\ref{H}). The spin-projection
$\sigma$ momentum distribution operator in Eq. (\ref{Popel}) is given by
$\hat{N}_{\sigma} (k) = c_{k,\,\sigma }^{\dagger }\,c_{k,\,\sigma }$. The operators
$c_{j,\,\sigma }^{\dagger }$ and $c_{j,\,\sigma}$ (and $c_{k,\,\sigma }^{\dagger }$ and
$c_{k,\,\sigma }$) which appear in the expressions of the above operators are the
spin-projection $\sigma $ electron creation and annihilation operators at site $j$ (and
carrying momentum $k$), respectively. Moreover, on the right-hand side of Eq. (\ref{H}),
$\mu_c=2\mu$, $\mu_s=2\mu_0 H$, $\mu_0$ is the Bohr magneton, and ${\hat{S }}_c^z=
-{1\over 2}[N_a-\hat{N}]$ and ${\hat{S }}_s^z= -{1\over
2}[{\hat{N}}_{\uparrow}-{\hat{N}}_{\downarrow}]$ are the diagonal generators of the
$\eta$-spin and spin $SU(2)$ algebras \cite{HL,Yang89}, respectively. We denote the
corresponding $\eta$-spin and spin state eigenvalues by $S_c$ and $S_s$, respectively.
The Hamiltonian $\hat{H}_{SO(4)}$ defined in Eq. (\ref{H}) commutes with the six
generators of these algebras. The off-diagonal generators are given in Eqs. (7) and (8)
of Ref. \cite{I}. We note that the Bethe-ansatz solution of the model refers to the
Hilbert subspace spanned by the lowest-weight states (LWSs) of the $\eta$-spin and spin
algebras, {\it i.e.} such that $S_{\alpha}= -S^z_{\alpha}$ where $\alpha =c,\,s$
\cite{HL,Yang89,I}.

The rotated-electron operator ${\tilde{c}}_{j,\,\sigma}^{\dag}$ is related to the
corresponding electronic operator $c_{j,\,\sigma}^{\dag}$ in Eq. (19) of Ref. \cite{I}.
This relation involves the electron - rotated-electron unitary operator ${\hat{V}}(U/t)$,
which is uniquely defined by Eqs. (21)-(23) of the same paper. (Expressions for the
unitary operator ${\hat{V}}(U/t)$ in terms of electronic elementary operators order by
order in $t/U$ are provided in Ref. \cite{Harris}.) The electron and rotated-electron
operators $c_{j,\,\sigma}^{\dag}$ and ${\tilde{c}}_{j,\,\sigma}^{\dag}$, respectively,
are only identical in the $U/t\rightarrow\infty$ limit where electron double occupation
becomes a good quantum number. The rotated-electron double occupation operator is given
in Eq. (20) of Ref. \cite{I}. Operators which commute with the electron -
rotated-electron unitary operator ${\hat{V}}(U/t)$ have the same expressions in terms of
both elementary electronic operators $c_{j,\,\sigma}^{\dag}$ and $c_{j,\,\sigma}$ and
rotated-electron operators ${\tilde{c}}_{j,\,\sigma}^{\dag}$ and
${\tilde{c}}_{j,\,\sigma}$. Examples of such operators are the six generators of the
$\eta$-spin and spin  $SU(2)$ algebras and the momentum operator (\ref{Popel}) \cite{I}.

The holons, spinons, and pseudoparticles studied in Ref. \cite{I} emerge from the
electron - rotated-electron unitary transformation which is such that rotated-electron
double occupation is a good quantum number. The holons have $\eta$ spin $1/2$ and spin
zero, whereas the spinons have spin $1/2$ and no $\eta$-spin degrees of freedom.
Throughout this paper we denote the holons and spinons according to their $\pm 1/2$
$\eta$-spin and spin projections, respectively. For the description of the transport of
charge in terms of electrons and electronic holes the $-1/2$ and $+1/2$ holons carry
charge $-2e$ and $+2e$, respectively. Based on symmetry considerations \cite{I}, we can
classify the $\pm 1/2$ holons and $\pm 1/2$ spinons into two classes: those which remain
invariant under the electron - rotated-electron unitary transformation, and those which
do not. The former are called $\pm 1/2$ Yang holons and $\pm 1/2$ HL spinons, with
numbers reading $L_{c,\,\pm 1/2}=[S_c\mp S_c^z]$ and $L_{s,\,\pm 1/2}=[S_s\mp S_s^z]$,
respectively, for all energy eigenstates. The latter are part of $\eta$-spin-zero
$2\nu$-holon composite $c\nu$ pseudoparticles and spin-zero $2\nu$-spinon composite
$s\nu$ pseudoparticles, respectively, where $\nu=1,2,...$ is the number of $+1/2$ and
$-1/2$ holon or $+1/2$ and $-1/2$ spinon pairs. Thus, the total number of $\pm 1/2$
holons $(\alpha =c)$ and $\pm 1/2$ spinons $(\alpha =s)$ reads $M_{\alpha,\,\pm
1/2}=L_{\alpha,\,\pm 1/2}+\sum_{\nu =1}^{\infty}\nu\,N_{\alpha\nu}$, where
$N_{\alpha\nu}$ denotes the number of composite $\alpha\nu$ pseudoparticles. The total
number of holons $(\alpha =c)$ and spinons $(\alpha =s)$ is then given by
$M_{\alpha}=L_{\alpha}+2\sum_{\nu =1}^{\infty}\nu\,N_{\alpha\nu}$ where
$L_{\alpha}=2S_{\alpha}$ denotes the total number of Yang holons $(\alpha =c)$ and HL
spinons $(\alpha =s)$. These numbers are such that $M_c=[N_a-N_c]$ and $M_s=N_c$, where
$N_c$ is the number of rotated-electron singly occupied sites and thus of $c$
pseudoparticles and $[N_a-N_c]$ the number of rotated-electron doubly occupied plus
unoccupied sites and of $c$ pseudoparticle holes \cite{I}. (In the above designations
{\it HL spinon} and {\it Yang holon}, HL stands for Heilmann and Lieb and Yang refers to
C. N. Yang, respectively, who are the authors of Refs. \cite{HL,Yang89}.) An useful
concept introduced in Ref. \cite{II} is that of a CPHS ensemble subspace where CPHS
stands for $c$ pseudoparticle, holon, and spinon. This is a Hilbert subspace spanned by
all energy eigenstates with fixed values for the $-1/2$ Yang holon number $L_{c,\,-1/2}$,
$-1/2$ HL spinon number $L_{s,\,-1/2}$, $c$ pseudoparticle number $N_c$, and for the sets
of $\alpha\nu$ pseudoparticle numbers $\{N_{c\nu}\}$ and $\{N_{s\nu}\}$ corresponding to
the $\nu=1,2,3,...$ branches.

It is found in later sections that there is a local $c$ pseudoparticle associated with
the bare-momentum $c$ pseudoparticle defined in terms of the Bethe-ansatz quantum numbers
in Ref. \cite{I}. The former pseudoparticle is a composite quantum object of a chargeon
and an antichargeon. The chargeon and the antichargeon describe the charge degrees of
freedom of the rotated electron and rotated-electronic hole of each singly occupied site
\cite{I}, respectively. It is also found that the local $c\nu$ pseudoparticle (and $s\nu$
pseudoparticle) description introduced in this paper is related by a Fourier transform to
the bare-momentum $c\nu$ pseudoparticle (and $s\nu$ pseudoparticle) representation
defined in terms of the Bethe-ansatz quantum numbers in Ref. \cite{I}. The former local
and composite $2\nu$-holon (and $2\nu$-spinon) quantum objects introduced here involve
the rotated electrons and rotated electronic holes associated with $\nu$ doubly occupied
sites and $\nu$ unoccupied sites (and $\nu$ spin-down and $\nu$ spin-up singly occupied
sites).

Finally, the application of the holon, spinon, and $c$ pseudoparticle description to the
evaluation of the finite-energy spectral-weight distributions \cite{V} involves a second
unitary transformation, which maps the $c$ pseudoparticles (and composite $c\nu$ or
$s\nu$ pseudoparticles) onto $c$ pseudofermions (and composite $c\nu$ or $s\nu$
pseudofermions) and is defined in the {\it pseudofermion subspace} (PS) \cite{IIIb}. Such
a transformation introduces shifts of order $1/L$ in the pseudoparticle discrete momentum
values and leaves all other pseudoparticle properties invariant. As a result of such
momentum shifts and in contrast to the $c$ pseudoparticles and composite $c\nu$ or $s\nu$
pseudoparticles, the corresponding pseudofermions have no residual-interaction energy
terms. However, in the PS where the one- and two-electron excitations are contained
\cite{V,IIIb}, the local $c$ pseudoparticle and composite $c\nu$ or $s\nu$ pseudoparticle
site distribution configurations that describe the energy eigenstates are the same as
those of the corresponding $c$ pseudofermions and composite $c\nu$ or $s\nu$
pseudofermions. Thus, since these site distribution configurations remain invariant under
the pseudoparticle - pseudofermion unitary transformation, our results apply both to
pseudoparticles and pseudofermions. In the studies of Secs. III - V we use the
pseudoparticle representation, once it is more general and refers to the whole Hilbert
space of the model (\ref{H}). As discussed in Sec. VI, both the results obtained in Secs.
III - V and the direct relation between pseudoparticles and pseudofermions provide useful
information for the correct identification of the scatterers and scattering centers
\cite{S} that control the model unusual spectral properties \cite{V}. For other aspects
of the holon, spinon, and pseudoparticle/pseudofermion description which are useful for
the studies of this paper see Refs. \cite{I,S,V}.

\section{ENERGY EIGENSTATES AND BASIC PROPERTIES ASSOCIATED WITH THE
PSEUDOPARTICLE INTERNAL STRUCTURE AND {\it UNOCCUPIED SITES}}

The studies of Ref. \cite{I} reveal that the energy eigenstates of the model are
described by the the same rotated-electron site distribution configurations for all
values of $U/t$ but do not provide such configurations. Since the electron -
rotated-electron unitary transformation becomes the unit transformation as
$U/t\rightarrow\infty$, in this section we study the $U/t$ independent rotated-electron
site distribution configurations which describe these states by considering the
corresponding electron site distribution configurations for the model in the limit
$U/t\rightarrow\infty$.

In such a limit there is a huge degeneracy of $\eta$-spin and spin occupancy
configurations. Such a degeneracy results from the simple form that the kinetic energy
$T$ and potential energy $V$ spectra associated with the operators (5) and (6) of Ref.
\cite{I}, respectively, have in the limit $U/t\rightarrow\infty$,

\begin{equation}
E_H = T + V \, ; \hspace{1cm} T = 2t \sum_{j=1}^{N_a}N_c (q_j)[-2t\cos q_j] \, ;
\hspace{1cm} V/U = D\, . \label{EHUinf}
\end{equation}
Here $E_H$ is the energy spectrum of the Hamiltonian ${\hat{H}}_H$ of Eq. (\ref{H}) and
the electron double occupation $D$ is a good quantum number which equals rotated-electron
double occupation in that limit. The kinetic energy equals that of a system of free
spin-less fermions \cite{Penc96-97,Ogata}. The corresponding momentum spectrum is given
in Eq. (36) of Ref. \cite{I}, with $M_{c,\,-1/2}$ replaced by electron double occupation
$D$. Since the energy spectrum (\ref{EHUinf}) is independent of the $\eta$-spin and spin
occupancy configurations, there are several choices for complete sets of energy
eigenstates with the same energy and momentum spectra. However, only one of these choices
is associated with the rotated-electron site distribution configurations which describe
the energy eigenstates for all values of $U/t$. A short discussion of the relation of our
results to some of the well known concepts of the $U/t\rightarrow\infty$ physics is
presented in the Appendix.

The pseudoparticle bare momentum obeys well defined boundary conditions which are a
necessary condition for the fulfilment of the periodic boundary conditions for the
original electrons. However, such a pseudoparticle bare-momentum boundary conditions are
not a sufficient condition to ensure the electronic periodic boundary conditions. A
second condition imposes that the internal structure of the local composite $\alpha\nu$
pseudoparticles introduced in later sections must be of a specific form. In this section
we introduce a set of properties which are rather useful for the following two issues:
First, for the construction of the specific rotated-electron site distribution
configurations which describe the internal structure of such local composite $\alpha\nu$
pseudoparticles; Second, for the definition of the corresponding pseudoparticle {\it
unoccupied sites} in terms of occupancy configurations of rotated-electron sites.

Our suitable basis choice corresponds to the bare-momentum energy eigenstates associated
with the thermodynamic Bethe-ansatz equations introduced by Takahashi \cite{Takahashi,I}.
For $U/t\rightarrow\infty$ there are other choices for complete sets of energy
eigenstates \cite{Geb}. Both the bare-momentum energy eigenstates and the {\it
symmetrized energy eigenstates} used in Ref. \cite{Geb} are superpositions of charge (and
spin) sequences formed by local electron distribution configurations of doubly occupied
and unoccupied sites (and spin-down and spin-up singly occupied sites). The expression of
both these two sets of energy eigenstates ensures the periodic boundary conditions for
the original electronic problem. However, only the former complete set of energy
eigenstates is of interest for the study of the finite $U/t$ problem studied here.

Throughout this paper we denote the rotated-electron doubly occupied and unoccupied sites
by $\bullet$ and $\circ$ (and the spin-down and spin-up rotated-electron singly occupied
sites by $\downarrow$ and $\uparrow$), respectively. In the $U/t\rightarrow\infty$ limit
such a concept also refers to electrons. Often we add an index to these symbols which
defines the position of the doubly occupied site or unoccupied  site (and spin-down
singly occupied site or spin-up singly occupied site). For the symmetrized energy
eigenstates the charge and spin sequences are properly symmetrized owing to the periodic
boundary conditions of the original problem. (This justifies the designation of {\it
symmetrized energy eigenstates}.) The procedure used in such a symmetrization involves
powers of suitable charge and spin operators ${\hat{\cal{T}}}_C$ and ${\hat{\cal{T}}}_S$,
respectively, for cyclic permutation of the electron site distribution configurations of
the local charge and spin sequences \cite{Geb}. For instance, let

\begin{equation}
(\bullet,\,\bullet,\,\circ,\,\bullet,\,\circ,...,\circ,\,\bullet,\,\bullet) \, ,
\label{cseq}
\end{equation}
and

\begin{equation}
(\downarrow,\,\uparrow,\,\downarrow,\,\uparrow,\,\uparrow,...,
\downarrow,\,\downarrow,\,\uparrow) \, , \label{sseq}
\end{equation}
be a charge and a spin sequence, respectively. Then the operators ${\hat{\cal{T}}}_C$ and
${\hat{\cal{T}}}_S$ are such that,

\begin{equation}
{\hat{\cal{T}}}_C\,(\bullet,\,\bullet,\,\circ,\,\bullet,\,\circ,...,\circ,
\,\bullet,\,\circ) =
(\circ,\,\bullet,\,\bullet,\,\circ,\,\bullet,\,\circ,...,\circ,\,\bullet)\, ,
\label{Tcseq}
\end{equation}
and

\begin{equation}
{\hat{\cal{T}}}_S\,(\downarrow,\,\uparrow,\,\downarrow,\,\uparrow,\,
\uparrow,...,\downarrow,\,\downarrow,\,\uparrow) =
(\uparrow,\,\downarrow,\,\uparrow,\,\downarrow,\,\uparrow,\,\uparrow,...,
\downarrow,\,\downarrow)\, , \label{Tsseq}
\end{equation}
respectively. The powers $[{\hat{\cal{T}}}_C]^{K_C}$ and $[{\hat{\cal{T}}}_S]^{K_S}$ were
called in Ref. \cite{Geb} $K_C$ and $K_S$, respectively. This introduces the {\it charge
momentum} $k_C$ and {\it spin momentum} $k_S$, respectively, such that,

\begin{equation}
k_C = {2\pi\over K_C\,a}\,m_C \, ; \hspace{0.25cm} m_C = 0,1,...,K_C-1 \, ;
\hspace{0.5cm} k_S = {2\pi\over K_S\,a}\,m_S \, ; \hspace{0.25cm} m_S = 0,1,...,K_S-1 \,
. \label{kCkS}
\end{equation}

The symmetrized energy eigenstates are classified according to their charge and spin
sequence, their charge momentum $k_C$ and spin momentum $k_S$, and a number $N_C$ of
momenta which in the notation of Ref. \cite{Geb} equals the number of charges. The latter
discrete momenta are closely related to the discrete bare-momentum values occupied by $c$
pseudoparticle holes in the pseudoparticle representation of the bare-momentum energy
eigenstates considered in Ref. \cite{I}. Moreover, the number $N_C$ of charges equals
both the number $N^h_c=[N_a-N_c]$ of $c$ pseudoparticle holes and the number $M_c$ of
holons. Thus, this is a good quantum number for both the bare-momentum and symmetrized
energy eigenstates. However, the charge momentum $k_C$ and spin momentum $k_S$ are
eigenvalues of the charge momentum operator ${\hat{k}}_C$ and spin momentum operator
${\hat{k}}_S$, respectively, which in general do not commute with the set of operators
$\{{\hat{N}}_{\alpha\nu}(q)\}$ of the bare-momentum basis considered in Ref. \cite{I}.
Thus, in general the bare-momentum energy eigenstates are not eigenstates of the charge
momentum operator ${\hat{k}}_C$ and spin momentum operator ${\hat{k}}_S$. Exceptions are
the bare-momentum energy eigenstates with occupancy of a single $c\nu$ pseudoparticle
(and a single $s\nu$ pseudoparticle) and with no finite occupancy of Yang holons (and HL
spinons) and of $c\nu'$ pseudoparticles (and $s\nu'$ pseudoparticles) belonging to other
branches such that $\nu'\neq\nu$. We find below that in this case the corresponding
bare-momentum energy eigenstate is also an eigenstate of the charge (and spin) momentum
operator of eigenvalue $k_C =\pi /a$ (and $k_S =\pi /a$).

According to Eqs. (B.1) and (B.2) of Ref. \cite{I}, the discrete values of the bare
momentum $q_j$ are such that $q_{j+1}-q_j = 2\pi/L$ and $q_j=[2\pi/L]\,I^c_j$ or
$q_j=[2\pi/L]\,I^{\alpha\nu}_j$ where the numbers $I^c_j$ and $I^{\alpha\nu}_j$ with
$j=1,2,...,N_a$ and $j=1,2,...,N^*_{\alpha\nu}$, respectively, are integers or half-odd
integers as a result of the following boundary conditions,

\begin{equation}
e^{iq_j\,L}=(e^{i\pi})^{\textstyle [
\sum_{\alpha=c,\,s}\sum_{\nu=1}^{\infty}N_{\alpha\nu}]} \, , \label{pbccp}
\end{equation}
in the case of the $c$ pseudoparticle branch and,

\begin{equation}
e^{iq_j\,L}=(e^{i\pi})^{[1+N^*_{\alpha\nu}]}=(e^{i\pi})^{[1+L_{\alpha}+
N_{\alpha\nu}]}=(e^{i\pi})^{[1+N_c + N_{\alpha\nu}]} \, ; \hspace{1cm} \alpha =c,s \, ,
\hspace{0.5cm} \nu =1,2,... \, , \label{pbcanp}
\end{equation}
for the $\alpha\nu$ pseudoparticle branches. Here
$N^*_{\alpha\nu}=N_{\alpha\nu}+N^h_{\alpha\nu}$ where the number $N^h_{\alpha\nu}$ is
defined by Eqs. (B.7) and (B.11) of Ref. \cite{I}. According to Eq. (B.14) of the same
paper, for the $\alpha\nu$ pseudoparticle branches the $j=1$ minimum and
$j=N^*_{\alpha\nu}$ maximum index of the bare-momentum values $q_j$ are such that,

\begin{equation}
-q_1 = q_{N^*_{\alpha\nu}} = q_{\alpha\nu} = {\pi\over L}[N^*_{\alpha\nu}-1] \, .
\label{limits}
\end{equation}
For the $c$ pseudoparticles the limiting values are $q_1=q_c^{-}$ and $q_{N_a}=q_c^{+}$
where the bare momenta $q_c^{\pm}$ are defined in Eqs. (B.15)-(B.17) of Ref. \cite{I}.

There is a holon, spinon, $c$ pseudoparticle separation for the whole parameter space of
the 1D Hubbard model \cite{I}. For the $t/U\rightarrow 0$ limit the description of the
$c$ pseudoparticle excitation sector is very similar for both the representations in
terms of bare-momentum energy eigenstates and symmetrized energy eigenstates. The
separation of the charge and spin excitation sectors occurs for these two alternative
representations. The construction of the bare-momentum energy eigenstates also involves
superpositions of charge (and spin) sequences associated with rotated-electron
distribution configurations of doubly occupied and unoccupied sites (and spin-down and
spin-up singly occupied sites). Such superpositions are also due to the periodic boundary
conditions but are not in general generated by the above symmetrization procedure. In
spite of these similarities, the complete sets of symmetrized and bare-momentum energy
eigenstates of the 1D Hubbard model in the limit of $t/U\rightarrow 0$ correspond in
general to different states. The difference between the symmetrized and bare-momentum
energy eigenstates refers in general both to the form of the local electronic
distribution configurations of doubly occupied and unoccupied sites (and spin-down and
spin-up singly occupied sites) which describe the charge (and spin) sequences and to the
form of the superposition of these local sequences which describes the energy
eigenstates.

It is useful for the introduction of the concepts of a local pseudoparticle and an
effective pseudoparticle lattice to consider the eight properties given below. However,
the precise definition of these concepts involves clarification of several other issues
beyond these properties and is only fulfilled in Sec. V. These properties are used in the
ensuing section in finding the rotated-electron site distribution configurations of the
local charge and spin sequences whose Fourier-transform superpositions describe the
bare-momentum energy eigenstates of the model. They follow in part from symmetries and
features of the pseudoparticle, holon, and spin description and related rotated-electron
representation studied in Refs. \cite{I,V,II} and from well known properties associated
with the $t/U\rightarrow 0$ physics. Although some of the results used in the following
properties are specific to the $t/U\rightarrow 0$ physics, we emphasize that some of the
conclusions also apply to finite values of $t/U$ provided that electrons are replaced by
rotated electrons. These useful properties read:\vspace{0.5cm}

1-III The numbers of electron doubly occupied sites, unoccupied sites, spin-down singly
occupied sites, and spin-up singly occupied sites are good quantum numbers whose values
are equal to the total numbers of $-1/2$ holons, $+1/2$ holons, $-1/2$ spinons, and
$+1/2$ spinons, respectively, of the bare-momentum energy eigenstates.\vspace{0.5cm}

2-III The kinetic energy $T$ given in Eq. (\ref{EHUinf}) arises from the movements of the
singly occupied sites relative to the doubly occupied and unoccupied  sides which do not
change double occupation. These movements are fully described by the $c$ pseudoparticles
which are associated with the charge degrees of freedom of these sites. In the limit
$t/U\rightarrow 0$ these quantum objects acquire a spin-less fermion spectrum. On the
other hand, the electron distribution configurations of doubly occupied and unoccupied
sites and of spin-down and spin-up singly occupied sites do not contribute to the kinetic
energy $T$. Moreover, these electron site distribution configurations must remain
unchanged in spite of the movements of the $c$ pseudoparticles. Alternatively, we can
consider that the electron doubly occupied and unoccupied sites move relative to the
singly occupied sites. In this case one describes the movements of the $c$
pseudoparticles in terms of the movements of the corresponding $c$ pseudoparticle holes.
This is the choice of Ref. \cite{Geb} for the case of the symmetrized energy eigenstates.
We recall that the numbers $N_c$ of $c$ pseudoparticles, $N^h_c$ of $c$ pseudoparticle
holes, $M_s$ of spinons, and $M_c$ of holons are such that $N_c+N^h_c=N_a$, $N_c =M_s$,
and $N^h_c=M_c$ and thus these two alternative descriptions are fully equivalent.
\vspace{0.5cm}

3-III We call {\it local charge sequences} and {\it local spin sequences} the occupancy
configurations of the $\pm 1/2$ holons (and corresponding electron distribution
configurations of doubly occupied and unoccupied sites) and the occupancy configurations
of the $\pm 1/2$ spinons (and corresponding electron distribution configurations of
singly occupied sites of spin projection $\pm 1/2$), respectively. These local charge
(and spin) sequences can also be expressed in terms of occupancy configurations of $\pm
1/2$ Yang holons and local $c\nu$ pseudoparticles (and $\pm 1/2$ HL spinons and local
$s\nu$ pseudoparticles) belonging to $\nu=1,2,3...$ branches. From the general properties
introduced in Ref. \cite{I}, one finds that the electronic description of these quantum
objects in terms of distribution configurations of electronic doubly occupied and
unoccupied sites (and spin-down and spin-up singly occupied sites) is as follows: A
$-1/2$ Yang holon (and $-1/2$ HL spinon) is described by a doubly occupied site (and a
spin-down singly occupied site); A $+1/2$ Yang holon (and $+1/2$ HL spinon) is described
by an unoccupied  site (and a spin-up singly occupied site); A local $c\nu$
pseudoparticle (and a local $s\nu$ pseudoparticle) is described by a number $\nu$ of
doubly occupied sites and a number $\nu$ of unoccupied sites (and a number $\nu$ of
spin-down singly occupied sites and a number $\nu$ of spin-up singly occupied sites). The
electron distribution configurations of doubly occupied and unoccupied sites (and of
spin-down and spin-up singly occupied sites) of any local charge (and spin) sequence can
be expressed in terms of a corresponding occupancy configuration of $\pm 1/2$ Yang holons
and local $c\nu$ pseudoparticles (and $\pm 1/2$ HL spinons and local $s\nu$
pseudoparticles) belonging to the $\nu=1,2,3...$ branches. The form of Eq. (36) of Ref.
\cite{I} reveals that a $-1/2$ holon carries momentum $\pi/a$ and that in the limit
$t/U\rightarrow 0$ each doubly occupied site must also be associated with a momentum
$\pi/a$, respectively. Therefore, the expressions of the local charge sequences must
include a phase factor operator $\exp (i\pi\sum_{j}j\,{\hat{D}}_j)$ such that the $j$
summation runs over the sites doubly occupied and unoccupied  and the local double
occupation operator has eigenvalues $1$ and $0$ if the site $j$ is doubly occupied and
unoccupied, respectively.\vspace{0.5cm}

4-III The number of different possible spatial discrete positions of a local $\alpha\nu$
pseudoparticle equals the number $N^*_{\alpha\nu}=N_{\alpha\nu}+N^h_{\alpha\nu}$, with
$N^h_{\alpha\nu}$ given in Eqs. (B.7) and (B.11) of Ref. \cite{I}. The number
$N^*_{\alpha\nu}$ is directly provided by the Bethe-ansatz solution \cite{I} since it
also equals the number of different discrete bare momentum values $q_j$, where
$j=1,2,...,N^*_{\alpha\nu}$, of the bare-momentum $\alpha\nu$ pseudoparticle band. Such a
local $\alpha\nu$ pseudoparticle has an internal structure corresponding to the electron
distribution configurations of $\nu$ doubly occupied sites and $\nu$ unoccupied sites
($\alpha =c$) or of $\nu$ spin-down and $\nu$ spin-up singly occupied sites ($\alpha
=s)$. Thus, each of the $N_{\alpha\nu}$ occupied locations does not correspond a single
lattice site but instead involves $2\nu$ lattice sites.\vspace{0.5cm}

5-III For energy eigenstates with finite occupancy of $-1/2$ and $+1/2$ Yang holons (and
$-1/2$ and $+1/2$ HL spinons) the electron distribution configurations of doubly occupied
and unoccupied sites (and spin-down and spin-up singly occupied sites) describing the
local $c\nu$ pseudoparticle (and local $s\nu$ pseudoparticle) must remain unchanged under
the application of the off-diagonal generators of the $\eta$-spin $SU(2)$ algebra given
in Eqs. (7) of Ref. \cite{I} (and off-diagonal generators of the spin $SU(2)$ algebra
given in Eq. (8) of the same reference). Moreover, for states with no $+1/2$ or $-1/2$
Yang holons (and no $+1/2$ or $-1/2$ HL spinons) application of the operators
${\hat{S}}_c^{\dag}=\sum_{j}(-1)^j\, c_{j,\,\downarrow}^{\dag} c_{j,\,\uparrow}^{\dag}$
or ${\hat{S}}_c=\sum_{j}(-1)^j\, c_{j,\,\uparrow}c_{j,\,\downarrow}$ (and
${\hat{S}}_s^{\dag}= \sum_{j}c_{j,\,\downarrow}^{\dag}c_{j,\,\uparrow}$ or
${\hat{S}}_s=\sum_{j}c_{j,\,\uparrow}^{\dag} c_{j,\,\downarrow}$) onto these states must
give zero. These requirements result from the $\eta$-spin singlet character of the $c\nu$
pseudoparticles (and spin singlet character of the $s\nu$ pseudoparticles). On the other
hand, the transformations generated by application of these off-diagonal generators onto
the electron distribution configurations of doubly occupied and unoccupied sites (and
spin-down and spin-up singly occupied sites) describing the $\pm 1/2$ Yang holons (and
$\pm 1/2$ HL spinons) must be the ones defined by the $\eta$-spin (and spin)
algebra.\vspace{0.5cm}

6-III Let us consider a local charge (and spin) sequence with no Yang holons (and no HL
spinons) and consisting of $\nu$ electron doubly occupied sites and $\nu$ electron
unoccupied sites (and $\nu$ electron spin-down singly occupied sites and $\nu$ electron
spin-up singly occupied sites). If such a sequence describes a single local $c\nu$
pseudoparticle (and local $s\nu$ pseudoparticle) it is properly symmetrized in such way
that the distribution configurations associated with the internal structure of that
quantum object remain unchanged under cyclic permutations. Furthermore, property 5-III
imposes that a bare-momentum energy eigenstate with the above Yang holon and $c\nu$
pseudoparticle (and HL spinon and $s\nu$ pseudoparticle) numbers is an eigenstate of the
charge (and spin) momentum operator ${\hat{k}}_C$ (and ${\hat{k}}_S$) of eigenvalue $k_C
=\pi /a$ (and $k_S =\pi /a$). \vspace{0.5cm}

7-III The local $c\nu$ pseudoparticle (and a $s\nu$ pseudoparticle) is a quantum object
whose internal structure involves $\nu$ doubly occupied sites and $\nu$ unoccupied sites
(and $\nu$ spin-down singly occupied sites and $\nu$ spin-up singly occupied sites). The
electron distribution configurations of the $2\nu$ sites which describe the internal
structure a local $c\nu$ pseudoparticle (and $s\nu$ pseudoparticle) are the same for all
local charge ($\alpha =c$) or spin ($\alpha =s$) sequences involved in the description of
the $4^{N_a}$ bare-momentum energy eigenstates. Such a property follows from the
indiscernible character of local $\alpha\nu$ pseudoparticles with the same spatial
position but involved in occupancy configurations describing different local charge
($\alpha =c$) or spin ($\alpha =s$) sequences. This indiscernible character of the local
$\alpha\nu$ pseudoparticles results from the corresponding indiscernible character of the
$\alpha\nu$ pseudoparticles of bare-momentum $q_j$, which are indistinguishable quantum
objects.\vspace{0.5cm}

8-III Since in the limit $t/U\rightarrow 0$ there is nearest-neighbor hopping only and it
does not change double occupation, the charge, spin, $c$ pseudoparticle separation
studied in Ref. \cite{I} implies that both the local charge and spin sequences of the
energy eigenstates must be separately conserved. Moreover, the periodic boundary
conditions of the original electronic problem are ensured if both the pseudoparticle
bare-momentum discrete values obey Eqs. (\ref{pbccp}) and (\ref{pbcanp}) and the
requirements of the basic properties 6-III and 7-III are fulfilled. The bare-momentum
energy eigenstates are Fourier-transform superpositions of local charge sequences, spin
sequences, and $c$ pseudoparticle sequences corresponding to Slater determinants
involving the pseudoparticle bare momentum $q_j$ and the spatial coordinate $x_j$ of
these quantum objects. Both for the $c$ pseudoparticle branch and the $\alpha\nu$
pseudoparticle branches with finite occupancy in a given state, such spatial coordinate
is the conjugate of the bare momentum $q_j$ of the above Fourier transforms. The spatial
coordinate $x_j$ of the $c$ pseudoparticles (and $\alpha\nu$ pseudoparticles) is
associated with an effective pseudoparticle lattice of length $L$, number of lattice
sites $N_a$ (and $N^*_{\alpha\nu}$) and lattice constant $a$ (and
$a_{\alpha\nu}=L/N^*_{\alpha\nu}$). The possible values of these spatial coordinates are
$x_j=a\,j$ where $j=1,2,3,...,N_a$ (and $x_j=a_{\alpha\nu}\,j$ where
$j=1,2,3,...,N^*_{\alpha\nu}$). The effective pseudoparticle lattices arise because the
occupancy configurations of the different pseudoparticle branches are separately
conserved. This implies a complex behavior for all electrons that singly occupy lattice
sites, whose charge degrees of freedom correspond to the $c$ pseudoparticles and separate
from the spin degrees of freedom, which refer to the spinons associated with the $s\nu$
pseudoparticles and HL spinons.\vspace{0.5cm}

These eight basic properties play an important role in the mechanisms which in the
$t/U\rightarrow 0$ limit determine the choice of the electron site distribution
configurations of the local charge, spin, and $c$ pseudoparticle sequences whose
Fourier-transform superpositions describe the bare-momentum energy eigenstates. Property
1-III follows from well known properties of the model in the limit $t/U\rightarrow 0$.
Property 2-III is consistent with the finding of Ref. \cite{II} that in the limit
$t/U\rightarrow 0$ only the $c$ pseudoparticles move and {\it carry} kinetic energy,
whereas the $\pm 1/2$ holons and $\pm 1/2$ spinons correspond to unchanged occupancy
configurations in that limit. Such a property results from well established features of
the Bethe-ansatz solution \cite{Ogata,Penc96-97,II}. Property 3-III is a consequence of
the combination of property 1-III with the relation of $\pm 1/2$ Yang holons, $\pm 1/2$
HL spinons, and $\alpha\nu$ pseudoparticles to $\pm 1/2$ holons and $\pm 1/2$ spinons. In
property 4-III the number $N^*_{\alpha\nu}$ plays a central role. The expression of that
number is valid for all values of $U/t$ and is provided by the Bethe-ansatz solution.
This is consistent with the electronic site distribution configurations which describe
the local $\alpha\nu$ pseudoparticles in the limit of $t/U\rightarrow 0$ being the same
as the rotated-electron site distribution configurations which describe these local
$\alpha\nu$ pseudoparticles for finite values of $t/U$. Property 5-III follows from
general symmetries of the model which are also valid for all values of $U/t$. This basic
property is also consistent with the above equivalence of the electronic site
distribution configurations in the limit of $t/U\rightarrow 0$ and of the corresponding
rotated-electron configurations for finite values of $t/U$. Property 6-III is a
consequence of the periodic boundary conditions in the particular case when a local
charge or spin sequence corresponds to a single $c\nu$ or $s\nu$ pseudoparticle,
respectively. The property 7-III results from the indiscernible character of the
$\alpha\nu$ pseudoparticles. In the case of bare-momentum pseudoparticles such an
indiscernible character is implicit in the description of the bare-momentum energy
eigenstates in terms of pseudoparticle occupancy configurations \cite{I}. Finally,
property 8-III is related to the periodic boundary conditions of the original electronic
problem and to the existence in this limit of nearest-neighbor hopping only which does
not change the value of double occupation. Such a property is also related to the charge,
spin, $c$ pseudoparticle separation and associated complex behavior of the
singly-occupied-sites rotated electrons for all energy scales and for the whole parameter
space of the model. In addition, for the composite $\alpha\nu$ pseudoparticle branches,
we find in Sec. V that the emergence of the effective pseudoparticle lattices mentioned
in this property is related to the separation of the internal and translational degrees
of freedom of the local $\alpha\nu$ pseudoparticles.

Below we often refer to the rotated-electron site distribution configurations only.
However, whenever referring to rotated electrons we mean implicitly that in the limit
$t/U\rightarrow 0$ the electronic site distribution configurations which describe the
bare-momentum energy eigenstates are the same.

\section{THE PSEUDOPARTICLE INTERNAL STRUCTURE AND COMPLETE SET OF
LOCAL STATES IN TERMS OF CHARGE, SPIN, AND $c$ PSEUDOPARTICLE SEQUENCES}

In this section we find the rotated-electron site distribution configurations which
describe the local charge, spin, and $c$ pseudoparticle sequences. This requires the
study of the internal structure of the local $\alpha\nu$ pseudoparticles. In addition,
such an analysis requires the introduction of the concept of an effective pseudoparticle
lattice. The first step for the introduction of such a concept is the definition of the
spatial positions of the rotated-electron site distribution configurations which describe
both the local $c$ and $\alpha\nu$ pseudoparticles and their {\it unoccupied sites}. We
find that each specific site distribution configurations of a local charge, spin, and $c$
pseudoparticle sequence defines a local state. The set of all different possible such
states constitutes a complete set of local states. In the ensuing section we express the
energy eigenstates of the model as Fourier-transform superpositions of these local
states.

\subsection{THE LOCAL $\alpha\nu$ PSEUDOPARTICLE INTERNAL STRUCTURE}

We start by restricting our study to rotated-electron site distribution configurations
which describe the lowest-weight states of both the $\eta$-spin and spin $SU(2)$
algebras. These states have no $-1/2$ Yang holons and no $-1/2$ HL spinons. Thus, in this
case the number of Yang holons (and of HL spinons) is such that $L_c=L_{c,+1/2}$ (and
$L_s=L_{s,+1/2}$).

In the limit $U/t\rightarrow\infty$ the $N_c$ $c$ pseudoparticles behave as spin-less
fermions in a lattice of $N_a=N_c+N^h_c=M_s+M_c$ sites. Such a lattice is nothing but the
effective $c$ pseudoparticle lattice mentioned in property 8-III. As for the
bare-momentum $q_j$, such an effective $c$ pseudoparticle lattice and its site
coordinates $x_j$ remain the same for the whole parameter space of the model. Also the
$c$ pseudoparticle occupancy configurations of such a lattice describing a given energy
eigenstate are the same for all values of $U/t$. At a fixed value $N_c=M_s$ of the
numbers $N_c$ of $c$ pseudoparticles and $M_s$ of spinons the number of occupancy
configurations of the local $c$ pseudoparticles in such an effective lattice is given by,

\begin{equation}
{N_a\choose N_{c}} = {N_a!\over N_c!\,N^h_c!} = {N_a!\over M_s!\,M_c!}\, . \label{NcNhc}
\end{equation}
Each of these $c$ pseudoparticle occupancy configurations of the effective $c$
pseudoparticle lattice defines a {\it local $c$ pseudoparticle sequence}.

Let $x_{j_l}=a\,j_l$ where $l=1,2,..., N_c$ be the actual set of occupied coordinates of
the effective $c$ pseudoparticle lattice out of the available $x_j=a\,j$ coordinate sites
such that $j=1,2,..., N_a$. We note that these $N_a$ sites have the same coordinates and
lattice constant $a$ as the $N_a$ sites of the rotated-electron lattice. Moreover, the
set of coordinates $x_{j_l}=a\,j_l$ where $l=1,2,..., N_c$ corresponding to the occupied
sites of the effective $c$ pseudoparticle lattice are precisely the same as the
coordinates of the rotated-electron singly occupied sites. Thus, we can define each of
the occupancy configurations whose total number is given in Eq. (\ref{NcNhc}) by the
coordinates in units of the lattice constant $a$ of the rotated-electron singly occupied
sites,

\begin{equation}
(j_{1},\,j_2,...,j_{N_c}) \, . \label{lcp}
\end{equation}
The locations occupied by the rotated-electron doubly occupied and unoccupied sites are
then these left over by the rotated-electron singly occupied sites. Since the sequence
(\ref{lcp}) also gives the locations of the local $c$ pseudoparticles in their effective
lattice, the sites left over by these pseudoparticles define the locations of the local
$c$ pseudoparticle holes. Thus, the relation of the numbers $N_c$ and $N_c^h$ of local
$c$ pseudoparticles and local $c$ pseudoparticle holes, respectively, to the
rotated-electron site distribution configurations of a given energy eigenstate confirms
that there are $N_c$ rotated-electron singly occupied sites and $N^h_c=[N_a-N_c]$
rotated-electron doubly-occupied and unoccupied sites, as stated in Ref. \cite{I} . From
the use of the results of that reference we find that these numbers can be expressed as
follows,

\begin{equation}
N^h_c=N_a - N_c = L_c + 2\sum_{\nu =1}^{\infty} \nu \, N_{c\nu} \, ; \hspace{0.5cm} N_c =
L_s + 2\sum_{\nu =1}^{\infty} \nu \, N_{s\nu} \, . \label{LcsLWS}
\end{equation}

The local charge sequences (and spin sequences) introduced in property 3-III involve only
the rotated-electron doubly-occupied and unoccupied  site distribution configurations
(and spin-down and spin-up rotated-electron singly occupied site distribution
configurations) of $N^h_c=[N_a-N_c]$ sites (and $N_c$ sites) out of a total number $N_a$
of sites. As a result of the independent conservation of the charge and spin sequences,
the rotated-electron site distribution configurations of a charge (and spin) sequence is
for any energy eigenstate obtained simply by omitting the $N_c$ singly occupied sites
(and $N^h_c=[N_a-N_c]$ doubly-occupied and unoccupied sites). Thus, a local charge (and
spin) sequence corresponds to the rotated-electron occupancy configurations of a site
domain involving $N^h_c=[N_a-N_c]$ sites (and $N_c$ sites). It follows that the number of
sites of such a domain depends on the specific state under consideration. For the
ground-state numbers provided in Appendix C of Ref. \cite{I}, the number of sites of the
rotated-electron lattice which belong to the charge and spin sequences is $[N_a-N]$ and
$N$, respectively.

Also the effective $\alpha\nu$ pseudoparticle lattices mentioned in property 8-III and
introduced below, result from independent conservation laws associated with each
$\alpha\nu$ pseudoparticle branch such that $\alpha=c,\,s$ and $\nu=1,2,...$. These
effective $c\nu$ pseudoparticle (and $s\nu$ pseudoparticle) lattices are generated below
by omission of a number $\sum_{\nu'=1}^{\infty} \Bigl[\nu + \nu' - \vert\nu -
\nu'\vert\Bigl] N_{c\nu'}-N_{c\nu}$ (and $\sum_{\nu'=1}^{\infty} \Bigl[\nu + \nu' -
\vert\nu - \nu'\vert\Bigl] N_{s\nu'}-N_{s\nu}$) of sites out of the total number
$N^h_c=[N_a-N_c]$ (and $N_c$) of sites of the local charge sequence (and spin sequence).
Again, the number of sites of the local charge (and spin) sequence which contribute to an
effective $c\nu$ pseudoparticle (and $s\nu$ pseudoparticle) lattice depends on the
specific state under consideration. For instance, for a ground state corresponding to
densities in the ranges $0\leq na\leq 1$ and $0\leq ma\leq na$ the number of sites of the
local charge sequence (and spin sequence) which contribute to an effective $c\nu$
pseudoparticle (and $s\nu$ pseudoparticle) lattice is $[N_a-N]$ (and $N_{\uparrow}$ for
the $s1$ pseudoparticle branch and $[N_{\uparrow}-N_{\downarrow}]$ for the $s\nu$
pseudoparticle branches such that $\nu>1$) out of the $[N_a-N]$ sites (and
$N=[N_{\uparrow}+N_{\downarrow}]$ sites) of such a sequence. On the other hand, the
number of sites of the effective $c$ pseudoparticle lattice equals the number $N_a$ of
sites of the rotated-electron lattice and is the same for all energy eigenstates.

The expressions of Eq. (\ref{LcsLWS}) show that the value of the number $L_c$ of Yang
holons (and $L_s$ of HL spinons) of the local charge (and spin) sequence is uniquely
determined by the values of the numbers of $c$ pseudoparticles and $c\nu$ pseudoparticles
($c$ pseudoparticles and $s\nu$ pseudoparticles) of the same sequence. Thus, for given
number $N^h_c=[N_a+N_c]$ of rotated-electron doubly-occupied and unoccupied sites (and
$N_c$ of rotated-electron singly occupied sites) we can uniquely define the
rotated-electron site distribution configurations of the local charge (and spin) sequence
by providing the sites occupied by local $c\nu$ pseudoparticles (and $s\nu$
pseudoparticles) belonging to the branches $\nu=1,2,...$ with finite occupancy. The
charge-sequence (and spin-sequence) sites of the rotated-electron lattice left over by
these local pseudoparticles define the positions of the Yang holons (and HL spinons). It
is useful to introduce the charge-sequence site index $h$ and the spin-sequence site
index $l$ such that,

\begin{equation}
h=1,2,...,[N_a-N_c] \, ; \hspace{1cm} l=1,2,...,N_c \, . \label{hl}
\end{equation}
The ordering of the charge-sequence (and spin-sequence) index corresponds to the
spatial-position order from the left to the right-hand side of the corresponding site in
the rotated-electron lattice. The position in such a lattice of a spin-sequence site of
index $l$ and of a charge-sequence site of index $h$ are given by,

\begin{equation}
x_{j_{l}}=j_{l}\,a \, , \hspace{0.5cm} l=1,2,...,N_c \, ; \hspace{1cm} x_{j_{h}}=
j_{h}\,a \, , \hspace{0.5cm} h=1,2,...,[N_a-N_c] \, , \label{chcl}
\end{equation}
where $j_l$ are the indices of Eq. (\ref{lcp}) which define the position of the
rotated-electron singly occupied sites and $j_h$ are the indices which define the
position of the rotated-electron doubly occupied and unoccupied sites. The spatial
position of the latter sites corresponds to the sites left over by the rotated-electron
singly occupied sites.

>From both the above analysis and property 3-III we conclude that one can uniquely specify
a given rotated-electron site distribution configuration by providing the position of the
$N_c$ sites occupied by local $c$ pseudoparticles, $2\sum_{\nu=1}^{\infty}\nu\,N_{c\nu}$
sites occupied by local $c\nu$ pseudoparticles, and $2\sum_{\nu=1}^{\infty}\nu\,N_{s\nu}$
sites occupied by local $s\nu$ pseudoparticles. The $N_c$ sites occupied by $c$
pseudoparticles define the position of the rotated-electron singly occupied sites. The
$[N_a-N_c]$ rotated-electron doubly-occupied and unoccupied sites are the sites left over
by the rotated-electron singly occupied sites. The $L_c=2S_c$ sites occupied by Yang
holons and the $L_s=2S_s$ sites occupied by HL spinons are the sites left over in the
charge and spin sequences, respectively, by the sites occupied by local $c\nu$
pseudoparticles and local $s\nu$ pseudoparticles, respectively. As mentioned above, to
start with we consider that all Yang holons (and HL spinons) have $\eta$-spin projection
(and spin projection) $+1/2$ and thus correspond to rotated-electron unoccupied sites
(and rotated-electron spin-up singly occupied sites). The generalization to
rotated-electron site distribution occupancies associated with states containing both
$\pm 1/2$ Yang holons (and $\pm 1/2$ HL spinons) is straightforward and is introduced
later in this section.

The Yang holons (and HL spinons) have no internal structure and are the simplest of the
quantum objects which occupy the charge (and spin) sequence of the rotated-electron
lattice. According to property 1-III, the $+1/2$ Yang holons (and $+1/2$ HL spinons)
correspond to rotated-electron unoccupied sites (and spin-up rotated-electron singly
occupied sites) of these sequences. On the other hand, a local $\alpha\nu$ pseudoparticle
has internal structure and thus is a more involved quantum object than a Yang holon or a
HL spinon. This internal structure corresponds to the rotated-electron distribution
configurations of the $\nu$ doubly occupied sites and $\nu$ unoccupied sites ($\alpha
=c$) or $\nu$ spin-down singly occupied sites and $\nu$ spin-up singly occupied sites
($\alpha =s$) which according to property 3-III describe such a local $\alpha\nu$
pseudoparticle. If in the rotated-electron site distribution configurations which
describe the local $c\nu$ pseudoparticles we replace doubly occupied sites $\bullet$ and
unoccupied sites $\circ$ by spin-down singly occupied sites $\downarrow$ and spin-up
singly occupied sites $\uparrow$, respectively, and omit the phase factors generated by
the operator $\exp ({i\pi\sum_{j}j\,{\hat{D}}_j})$ mentioned in property 3-III, we obtain
the corresponding distribution configurations of the $s\nu$ pseudoparticles. Therefore,
often we consider the rotated-electron site distribution configurations which describe
the local $c\nu$ pseudoparticles only.

It is useful to classify the $2\nu$ lattice sites of the local charge sequence involved
in the description of a local $c\nu$ pseudoparticle into two sets of $\nu$ sites each.
Let us denote the index of these two sets of $\nu$ sites by $h_{j,\,g}$ and $h_{j,\,\nu
+g}$, respectively. Here $g=1,2,...,\nu$ refers to the internal site structure of the
local $c\nu$ pseudoparticle and the index $j=1,2,...,N^*_{c\nu}$ corresponds to its
position $[h_j\,a]$ in the rotated-electron lattice where,

\begin{equation}
h_j = {j_{h_{j,\,\nu}}+j_{h_{j,\,\nu +1}}\over 2} \, , \label{stringC}
\end{equation}
and the numbers $j_{h_{j,\,\nu}}$ are the indices $j_h$ of Eq. (\ref{lcp}) which define
the position of the rotated-electron doubly occupied/unoccupied sites in such a lattice.
Combination of the values of the indices $j=1,2,...,N^*_{c\nu}$ and $g=1,2,...,\nu$ fully
defines the position of the above $2\nu$ sites. The internal-structure index $g$ is such
that $h_{j,\,1}<h_{j,\,2}<...<h_{j,\,\nu}$ and $h_{j,\,\nu +1} <h_{j,\,\nu
+2}<...<h_{j,\,2\nu}$, respectively, where $h_{j,\,\nu} <h_{j,\,\nu +1}$. Equivalently,
often we denote these $2\nu$ internal-structure indices simply by $h_{j,\,x}$, where
$x=1,2,...,2\nu$ and $h_{j,\,1}<h_{j,\,2}<...<h_{j,\,2\nu}$. The set of $2\nu$
internal-structure indices $\{h_{j,\,1},\,h_{j,\,2},...,h_{j,\,2\nu}\}$ is in general a
sub-set of the $[N_a-N_c]$ charge-sequence indices $h$ given in Eq. (\ref{hl}). The
latter indices define the position of the charge-sequence rotated-electron doubly
occupied and unoccupied sites. We can also define the {\it charge-sequence position} of
the local $c\nu$ pseudoparticle which is defined as,

\begin{equation}
{\bar{h}}_j = {h_{j,\,\nu}+h_{j,\,\nu +1}-1\over 2} \, . \label{bstringC}
\end{equation}
The same definitions hold for the local $s\nu$ pseudoparticles with the indices
$h_{j,\,g}$ and $h_{j,\,\nu +g}$ replaced by the indices $l_{j,\,g}$ and $l_{j,\,\nu
+g}$, respectively, and thus the equivalent indices $h_{j,\,x}$ replaced by $l_{j,\,x}$.
The position of the local $s\nu$ pseudoparticle in the rotated-electron lattice is
defined as $[l_j\,a]$ where,

\begin{equation}
l_j = {j_{l_{j,\,\nu}}+j_{l_{j,\,\nu +1}}\over 2} \, , \label{stringS}
\end{equation}
$j=1,2,...,N^*_{s\nu}$, and $j_l$ are the indices of Eq. (\ref{lcp}) which define the
position of the rotated-electron singly occupied sites. On the other hand, the {\it
spin-sequence position} of the local $s\nu$ pseudoparticle is defined as,

\begin{equation}
{\bar{l}}_{j} = {l_{j,\,\nu}+l_{j,\,\nu +1}-1\over 2} \, , \label{bstringS}
\end{equation}
where again $j=1,2,...,N^*_{s\nu}$. We note that the indices ${\bar{h}}_j$ (and
${\bar{l}}_{j}$) given in Eq. (\ref{bstringC}) (and Eq. (\ref{bstringS})) which define
the charge-sequence (and spin-sequence) position of the local $c\nu$ pseudoparticle (and
local $s\nu$ pseudoparticle) are always positive integer numbers $1,\,2,\,3,...$.
According to Eq. (\ref{stringC}) (and Eq. (\ref{stringS})) the position $[h_j\,a]$ (and
$[l_j\,a]$) of the local $c\nu$ pseudoparticle (and local $s\nu$ pseudoparticle) in the
$N_a$-site rotated-electron lattice refers to a single point inside the $2\nu$-site
domain associated with such a quantum object. On the other hand, the charge-sequence
position (and spin-sequence position) of the local $c\nu$ pseudoparticles (and local
$s\nu$ pseudoparticles) defined by the index of Eq. (\ref{bstringC}) (and Eq.
(\ref{bstringS})) refers to the position of that quantum object relative to the
$[N_a-N_c]$ sites (and $N_c$ sites) of the charge (and spin) sequence only.

Below we clarify the following two issues: First, we find the rotated-electron
distribution configurations of the $2\nu$ sites which describe the internal structure of
a local $\alpha\nu$ pseudoparticle; Second, we find the rotated-electron site
distribution configurations which define the $N^h_{\alpha\nu}$ charge-sequence
($\alpha=c$) or spin-sequence ($\alpha=s$) {\it unoccupied sites} corresponding to the
local $\alpha\nu$ pseudoparticle branch. This study reveals that the position in the
rotated-electron lattice of the sites associated with the $N^h_{\alpha\nu}$
charge-sequence or spin-sequence {\it unoccupied sites} of each local $\alpha\nu$
pseudoparticle branch is uniquely determined by the position of the local $c$
pseudoparticles, local $c\nu$ pseudoparticles, and local $s\nu$ pseudoparticles. The same
holds for the positions of the Yang holons and HL spinons, as discussed above. Below we
also confirm that for fixed values of the numbers $N_{\alpha\nu}$ and $N^*_{\alpha\nu}$
the number of occupancy configurations of the local $\alpha\nu$ pseudoparticles is given
by,

\begin{equation}
{N_{\alpha\nu}^*\choose N_{\alpha\nu}} = {N_{\alpha\nu}^*!\over
N_{\alpha\nu}!\,N^h_{\alpha\nu}!}  \, . \label{NanNS}
\end{equation}
We classify each of the local $\alpha\nu$ pseudoparticle occupancy configurations by
providing the indices $h_j$ or $l_j$ given in Eqs. (\ref{stringC}) or (\ref{stringS}),
respectively, corresponding to the $N_{c\nu}$ and $N_{s\nu}$ pseudoparticle positions in
the rotated-electron lattice,

\begin{equation}
(h_{1},\,h_2,...,h_{N_{c\nu}}) \, , \label{localcnp}
\end{equation}
and

\begin{equation}
(l_{1},\,l_2,...,l_{N_{s\nu}}) \, , \label{localsnp}
\end{equation}
respectively. From the combination of Eqs. (\ref{NcNhc}) and (\ref{NanNS}), it follows
that the number of occupancy configurations of the local $c$ and $\alpha\nu$
pseudoparticles of a CPHS ensemble subspace is given by,

\begin{equation}
{\cal{N}}_{CPHS-es} = {N_a\choose N_c}\prod_{\alpha =c,s}\prod_{\nu =1}^{\infty}\,
{N_{\alpha\nu}^*\choose N_{\alpha\nu}} \, . \label{Ncphs}
\end{equation}

Let us denote each local state representing a specific local charge, spin, and $c$
pseudoparticle sequence whose number for a given CPHS ensemble subspace is given in Eq.
(\ref{Ncphs}) by,

\begin{equation}
\vert(j_{1},\,j_2,...,j_{N_c});\,
\{(h_{1},\,h_2,...,h_{N_{c\nu}})\};\,\{(l_{1},\,l_2,...,l_{N_{s\nu}})\}\rangle \, .
\label{localst}
\end{equation}
Here

\begin{equation}
\{(h_{1},\,h_2,...,h_{N_{c\nu}})\} = (h_1,\,h_2,...,h_{N_{c1}});\,
(h_{1},\,h_2,...,h_{N_{c2}});\,(h_{1},\,h_2,...,h_{N_{c3}});...\, , \label{argc}
\end{equation}
gives the positions in the rotated-electron lattice of the local $c\nu$ pseudoparticles
belonging to branches with finite occupancy in the state under consideration and,

\begin{equation}
\{(l_{1},\,l_2,...,l_{N_{s\nu}})\} = (l_{1},\,l_2,...,l_{N_{s1}});\,
(l_{1},\,l_2,...,l_{N_{s2}});\,(l_{1},\,l_2,...,l_{N_{s3}});...\, , \label{args}
\end{equation}
provides the positions in the same lattice of the local $s\nu$ pseudoparticles belonging
to branches with finite occupancy in the same state.

We emphasize that the number (\ref{Ncphs}) of such local states which equals the number
of different local pseudoparticle occupancy configurations of a CPHS ensemble subspace
indeed equals the dimension of such a subspace. It is straightforward to confirm from the
results of Ref. \cite{I} that the number of bare-momentum energy eigenstates that span
such a CPHS ensemble subspace also has the same value. This suggests that the set of
local states of form (\ref{localst}) is complete in that subspace, as is confirmed below.
We find later on in this section that the generalization of the present analysis to
rotated-electron site distribution configurations describing energy eigenstates with
finite occupancies of $\pm 1/2$ Yang holons and $\pm 1/2$ HL spinons requires the
introduction of the numbers $L_{c,\,-1/2}$ of $-1/2$ Yang holons and $L_{s,\,-1/2}$ of
$-1/2$ HL spinons in the labeling of the local states (\ref{localst}).

In order to find the rotated-electron site distribution configurations which describe the
internal structure of a local $\alpha\nu$ pseudoparticle, we use mainly the basic
properties 3-III, 5-III, 6-III, and 7-III. Often we consider the specific case of a local
$c\nu$ pseudoparticle. The internal structure of such a quantum object is studied by
considering first a charge sequence constituted by $\nu$ rotated-electron doubly occupied
sites and $\nu$ rotated-electron unoccupied sites. The use of properties 5-III and 6-III
leads to the finding of the specific rotated-electron site distribution configurations
which describe the local $c\nu$ pseudoparticle. Property 7-III then states that the
obtained rotated-electron site distribution configurations describe a local $c\nu$
pseudoparticle in any charge sequence.

According to property 3-III, the internal structure of a local $c\nu$ pseudoparticle
involves a number $\nu$ of rotated-electron doubly occupied sites and an equal number of
rotated-electron unoccupied sites. There is a number ${2\nu\choose \nu}$ of different
distribution configurations of these $\nu$ rotated-electron doubly occupied sites and
$\nu$ rotated-electron unoccupied sites. For simplicity, let us restrict the present
preliminary analysis to charge sequences with $\nu$ rotated-electron doubly occupied
sites, an equal number of rotated-electron unoccupied sites, and no Yang holons. Below we
find that a local $c\nu$ pseudoparticle is a superposition of a number $2^{\nu}$ of the
above rotated-electron site distribution configurations. For $\nu=1$ the local $c1$
pseudoparticle is a properly symmetrized superposition of the two available
rotated-electron site distribution configurations. For $\nu=2$ the local $c2$
pseudoparticle is a properly symmetrized superposition of four out of the six available
rotated-electron site distribution configurations. Thus, states including all possible
six different distribution configurations are a superposition of two states including two
local $c1$ pseudoparticles and a local $c2$ pseudoparticle, respectively. For $\nu=3$ the
local $c3$ pseudoparticle is a properly symmetrized superposition of eight out of the
twenty available rotated-electron site distribution configurations. States including all
possible twenty different distribution configurations are a superposition of three states
including (i) three local $c1$ pseudoparticles, (ii) a local $c1$ pseudoparticle and a
local $c2$ pseudoparticle, and (iii) a local $c3$ pseudoparticle. In the general case one
has that the local $c\nu$ pseudoparticle is a properly symmetrized superposition of
$2^{\nu}$ out of the ${2\nu\choose \nu}$ available rotated-electron site distribution
configurations. States including all possible ${2\nu\choose \nu}$ different distribution
configurations are a superposition of several states including different numbers
$N_{c\nu'}$ of $c\nu'$ pseudoparticles such that $\nu'=1,2,...,\nu$ and
$\sum_{\nu'=1}^{\nu}\nu'\,N_{c\nu'}=\nu$.

Among the $2^{\nu}$ rotated-electron site distribution configurations of a $c\nu$
pseudoparticle there is always a distribution configuration where the first $\nu$ sites
of index $h_{j,\,g}$ and $g=1,2,...,\nu$ are doubly occupied by rotated electrons and the
last $\nu$ sites of index $h_{j,\,\nu +g}$ and $g=1,2,...,\nu$ are free of rotated
electrons. We represent such a local $c\nu$ pseudoparticle rotated-electron site
distribution configuration of a charge sequence by,

\begin{equation}
(\bullet_{h_{j,\,1}},...,\bullet_{h_{j,\,\nu}}, \,\circ_{h_{j,\,1+\nu}},...,
\circ_{h_{j,\,2\nu}}) \, . \label{lcnp}
\end{equation}
The same applies to a local $s\nu$ pseudoparticle, the rotated-electron site distribution
configuration of a spin sequence corresponding to the one given in Eq. (\ref{lcnp})
being,

\begin{equation}
(\downarrow_{l_{j,\,1}},..., \downarrow_{l_{j,\,\nu}},\,\uparrow_{l_{j,\,1+\nu}},...,
\uparrow_{l_{j,\,2\nu}}) \, . \label{lsnp}
\end{equation}
We confirm below that the $c\nu$ pseudoparticle located in the rotated-electron lattice
at position $[h_j\,a]$ and the $s\nu$ pseudoparticle located in the same lattice at
position $[l_j\,a]$ are described by the following properly symmetrized superposition of
$2^{\nu}$ rotated-electron site distribution configurations,

\begin{equation}
\Bigl[\prod_{x=1}^{2\nu}e^{i\pi h_{j,\,x}{\hat{D}}_{j,\,x}}\Bigr]\,
\Bigl[\prod_{g=1}^{\nu}(1-{\hat{\cal{T}}}_{c\nu,\,j,\,g})\Bigr]
\,(\bullet_{h_{j,\,1}},...,\bullet_{h_{j,\,\nu}}, \,\circ_{h_{j,\,1+\nu}},...,
\circ_{h_{j,\,2\nu}}) \, ; \hspace{0.5cm} j=1,2,...,N_{c\nu} \, , \label{cnp}
\end{equation}
and

\begin{equation}
\Bigl[\prod_{g=1}^{\nu}(1-{\hat{\cal{T}}}_{s\nu,\,j,\,g})\Bigr] \,
(\downarrow_{l_{j,\,1}},..., \downarrow_{l_{j,\,\nu}},\,\uparrow_{l_{j,\,1+\nu}},...,
\uparrow_{l_{j,\,2\nu}}) \, ; \hspace{0.5cm} j=1,2,...,N_{s\nu} \, , \label{snp}
\end{equation}
respectively. According to property 3-III, the site-$h_{j,\,x}$ rotated-electron double
occupation operator ${\hat{D}}_{j,\,x}$ appearing here has eigenvalue $1$ and $0$ when
such a site is doubly occupied by rotated electrons and free of rotated electrons,
respectively, and the operator ${\hat{\cal{T}}}_{c\nu,\,j,\,g}$ (and
${\hat{\cal{T}}}_{s\nu,\,j,\,g}$) acts onto the pair of sites of indices $h_{j,\,g}$ and
$h_{j,\,g+\nu}$ (and $l_{j,\,g}$ and $l_{j,\,g+\nu}$) only. This operator always acts
onto rotated-electron site distribution configurations of the particular form illustrated
in Eq. (\ref{lcnp}) (and in Eq. (\ref{lsnp})). From the application of this operator onto
such a rotated-electron site distribution configuration, a new distribution configuration
is generated where the site of index $h_{j,\,g}$ is free of rotated electrons, the site
of index $h_{j,\,g+\nu}$ is doubly occupied by rotated electrons, and the occupancy of
the other $2(\nu-1)$ sites remains unchanged, {\it i.e.}

\begin{equation}
{\hat{\cal{T}}}_{c\nu,\,j,\,g}\,(...,\bullet_{h_{j,\,g}},...,\circ_{h_{j,\,g+\nu}},...) =
(...,\circ_{h_{j,\,g}},...,\bullet_{h_{j,\,g+\nu}},...) \, . \label{Tc}
\end{equation}
The same transformation law,

\begin{equation}
{\hat{\cal{T}}}_{s\nu,\,j,\,g}\,(...,\downarrow_{l_{j,\,g}},...,\uparrow_{l_{j,\,g+\nu}},...)
= (...,\uparrow_{l_{j,\,g}},...,\downarrow_{l_{j,\,g+\nu}},...) \, , \label{Ts}
\end{equation}
is associated with the application of the operator ${\hat{\cal{T}}}_{s\nu,\,j,\,g}$ onto
a rotated-electron site distribution configuration of the type illustrated in Eq.
(\ref{lsnp}).

The $2^{\nu}$ internal rotated-electron site distribution configurations on the
right-hand side of Eq. (\ref{cnp}) (and Eq. (\ref{snp})) are generated by considering
that in each of the $g=1,2,...,\nu$ pairs of sites of indices $h_{j\,g}$ and
$h_{j,\,g+\nu}$, the site of index $h_{j\,g}$ is doubly occupied by rotated electrons and
the site of index $h_{j,\,g+\nu}$ is free of rotated electrons and vice versa. In
general, we call $h_{j,\,x} \leftrightarrow h_{j,\,x'}$ {\it site pair} a superposition
of two rotated-electron distribution configurations of a rotated-electron doubly occupied
site and a rotated-electron unoccupied site where $h_{j,\,x}$ and $h_{j,\,x'}$, such that
$x,x'=1,...,2\nu$ and $x'> x$, are the indices of the two lattice sites involved. In the
first rotated-electron site distribution configuration, the sites of indices $h_{j,\,x}$
and $h_{j,\,x'}$ are doubly occupied by rotated electrons and free of rotated electrons,
respectively. In the second rotated-electron site distribution configuration, the sites
of indices $h_{j,\,x}$ and $h_{j,\,x'}$ are free of rotated electrons and doubly occupied
by rotated electrons, respectively. The rotated-electron distribution configurations of
all the remaining sites of the charge sequence except these two are identical. The

\begin{equation}
h_{j,\,x} \leftrightarrow h_{j,\,x'} \, ; \hspace{1cm} x,x'=1,...,2\nu \, ;
\hspace{0.5cm} x'>x \, , \label{sitep}
\end{equation}
site pair is defined as a superposition of these two rotated-electron site distributions
configurations where the first and the second distributions are multiplied by a phase
factor of $1$ and $-1$, respectively.

In figure 1 we represent a $h_{j,\,x} \leftrightarrow h_{j,\,x'}$ site pair by two
vertical lines connected by a horizontal line. The two vertical lines have the same
height and connect the corresponding site of the charge sequence to the horizontal line.
Thus, the two sites of a charge sequence connected by three such lines are assumed to be
involved in a superposition of two rotated-electron site distribution configurations
multiplied by the phase factors of $1$ and $-1$, respectively, where according to the
above recipe one of these sites is doubly occupied by rotated electrons and the other one
is free of rotated electrons and vice versa. Similarly, we call

\begin{equation}
l_{j,\,x} \leftrightarrow l_{j,\,x'} \, ; \hspace{1cm} x,x'=1,...,2\nu \, ;
\hspace{0.5cm} x'>x \, , \label{lsitep}
\end{equation}
site pair an equivalent superposition of two rotated-electron site distribution
configurations of a spin sequence. The definitions are the same provided that we replace
rotated-electron doubly occupied sites and rotated-electron unoccupied sites by spin-down
rotated-electron singly occupied sites and spin-up rotated-electron singly occupied
sites, respectively. Such a rotated-electron singly occupied site pair is also
graphically represented by the two vertical lines connected by a horizontal line plotted
in Fig. 1. The concept of a site pair plays an important role in the description of the
rotated-electron site distribution configurations of a local $\alpha\nu$ pseudoparticle.

\begin{figure*}
\includegraphics[width=7cm,height=5.00cm]{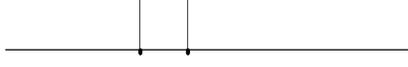}
\caption{\label{fig1} Graphical representation of a charge $h_{j,\,x} \leftrightarrow
h_{j,\,x'}$ site pair or of a spin $l_{j,\,x} \leftrightarrow l_{j,\,x'}$ site pair. In
the particular case of $x=1$ and $x'=2$ the figure represents the $h_{j,\,1}
\leftrightarrow h_{j,\,2}$ site pair of a $c1$ pseudoparticle or of the $l_{j,\,1}
\leftrightarrow l_{j,\,2}$ site pair of a $s1$ pseudoparticle.}
\end{figure*}

It is confirmed below that the rotated-electron site distribution configurations which
describe the internal structure of a local $\alpha\nu$ pseudoparticle always involve a
number $\nu$ of site pairs such that $x=g$ and $x'=g+\nu$ where $g=1,...,\nu$. Since
there are two possible rotated-electron site distribution configurations for each pair
and the number of pairs of each local $\alpha\nu$ pseudoparticle is $\nu$, the total
number of different internal rotated-electron site distribution configurations is indeed
$2^{\nu}$. For example, for a $c\nu$ pseudoparticle located in the rotated-electron
lattice at the position $[h_j\,a]$, the $2^{\nu}$ rotated-electron site distribution
configurations superposed in expression (\ref{cnp}) are the following,

\begin{equation}
\Bigl[\prod_{x=1}^{2}e^{i\pi
h_{j,\,x}{\hat{D}}_{j,\,x}}\Bigr]\,(1-{\hat{\cal{T}}}_{c1,\,j,\,g})
\,(\bullet_{h_{j,\,1}},\, \circ_{h_{j,\,2}}) = - e^{i\pi
h_{j,\,1}}\,(\bullet_{h_{j,\,1}},\, \circ_{h_{j,\,2}}) + e^{i\pi
h_{j,\,2}}\,(\circ_{h_{j,\,1}},\, \bullet_{h_{j,\,2}}) \, ,
\label{c1p}
\end{equation}
for the $c1$ pseudoparticle,

\begin{eqnarray}
& & \Bigl[\prod_{x=1}^{4}e^{i\pi h_{j,\,x}{\hat{D}}_{j,\,x}}\Bigr]
\,\Bigl[\prod_{g=1}^{2}(1-{\hat{\cal{T}}}_{c2,\,j,\,g})\Bigr]
\,(\bullet_{h_{j,\,1}},\,\bullet_{h_{j,\,2}},\,
\circ_{h_{j,\,3}},\,\circ_{h_{j,\,4}}) = \nonumber \\
& + & e^{i\pi [h_{j,\,1}+h_{j,\,2}]}\, (\bullet_{h_{j,\,1}},\,\bullet_{h_{j,\,2}},\,
\circ_{h_{j,\,3}},\,\circ_{h_{j,\,4}})
 - e^{i\pi[h_{j,\,2}+h_{j,\,3}]}
\,(\circ_{h_{j,\,1}},\,\bullet_{h_{j,\,2}},\,
\bullet_{h_{j,\,3}},\,\circ_{h_{j,\,4}}) \nonumber \\
& + & e^{i\pi [h_{j,\,3}+h_{j,\,4}]}\,(\circ_{h_{j,\,1}},\,\circ_{h_{j,\,2}},\,
\bullet_{h_{j,\,3}},\,\bullet_{h_{j,\,4}}) - e^{i\pi [h_{j,\,1}+h_{j,\,4}]}\,
(\bullet_{h_{j,\,1}},\,\circ_{h_{j,\,2}},\, \circ_{h_{j,\,3}},\,\bullet_{h_{j,\,4}}) \, ,
\label{c2p}
\end{eqnarray}
in the case of the $c2$ pseudoparticle, and

\begin{eqnarray}
& & \Bigl[\prod_{x=1}^{6}e^{i\pi h_{j,\,x}{\hat{D}}_{j,\,x}}\Bigr]
\,\Bigl[\prod_{g=1}^{3}(1-{\hat{\cal{T}}}_{c3,\,j,\,g})\Bigr]\,
(\bullet_{h_{j,\,1}},\,\bullet_{h_{j,\,2}},\, \bullet_{h_{j,\,3}},\,\circ_{h_{j,\,4}},\,
\circ_{h_{j,\,5}},\,\circ_{h_{j,\,6}}) = \nonumber \\
& - &
e^{i\pi[h_{j,\,1}+h_{j,\,2}+h_{j,\,3}]}\,(\bullet_{h_{j,\,1}},\,\bullet_{h_{j,\,2}},\,
\bullet_{h_{j,\,3}},\,\circ_{h_{j,\,4}},\,
\circ_{h_{j,\,5}},\,\circ_{h_{j,\,6}}) \nonumber \\
& + &
e^{i\pi[h_{j,\,2}+h_{j,\,3}+h_{j,\,4}]}\,(\circ_{h_{j,\,1}},\,\bullet_{h_{j,\,2}},\,
\bullet_{h_{j,\,3}},\,\bullet_{h_{j,\,4}},\,
\circ_{h_{j,\,5}},\,\circ_{h_{j,\,6}}) \nonumber \\
& - & e^{i\pi[h_{j,\,3}+h_{j,\,4}+h_{j,\,5}]}\,(\circ_{h_{j,\,1}},\,\circ_{h_{j,\,2}},\,
\bullet_{h_{j,\,3}},\,\bullet_{h_{j,\,4}},\,
\bullet_{h_{j,\,5}},\,\circ_{h_{j,\,6}}) \nonumber \\
& + &
e^{i\pi[h_{j,\,1}+h_{j,\,3}+h_{j,\,5}]}\,(\bullet_{h_{j,\,1}},\,\circ_{h_{j,\,2}},\,
\bullet_{h_{j,\,3}},\,\circ_{h_{j,\,4}},\,
\bullet_{h_{j,\,5}},\,\circ_{h_{j,\,6}}) \nonumber \\
& - &
e^{i\pi[h_{j,\,1}+h_{j,\,5}+h_{j,\,6}]}\,(\bullet_{h_{j,\,1}},\,\circ_{h_{j,\,2}},\,
\circ_{h_{j,\,3}},\,\circ_{h_{j,\,4}},\,
\bullet_{h_{j,\,5}},\,\bullet_{h_{j,\,6}}) \nonumber \\
& + &
e^{i\pi[h_{j,\,1}+h_{j,\,2}+h_{j,\,6}]}\,(\bullet_{h_{j,\,1}},\,\bullet_{h_{j,\,2}},\,
\circ_{h_{j,\,3}},\,\circ_{h_{j,\,4}},\,
\circ_{h_{j,\,5}},\,\bullet_{h_{j,\,6}}) \nonumber \\
& - &
e^{i\pi[h_{j,\,2}+h_{j,\,4}+h_{j,\,6}]}\,(\circ_{h_{j,\,1}},\,\bullet_{h_{j,\,2}},\,
\circ_{h_{j,\,3}},\,\bullet_{h_{j,\,4}},\,
\circ_{h_{j,\,5}},\,\bullet_{h_{j,\,6}}) \nonumber \\
& + & e^{i\pi[h_{j,\,4}+h_{j,\,5}+h_{j,\,6}]}\,(\circ_{h_{j,\,1}},\,\circ_{h_{j,\,2}},\,
\circ_{h_{j,\,3}},\,\bullet_{h_{j,\,4}},\, \bullet_{h_{j,\,5}},\,\bullet_{h_{j,\,6}}) \,
, \label{c3p}
\end{eqnarray}
for a $c3$ pseudoparticle. Similar expressions for local $s1$, $s2$, or $s3$
pseudoparticles are obtained by replacing in Eqs. (\ref{c1p})-(\ref{c3p}) the $\bullet$
and $\circ$ sites by $\downarrow$ and $\uparrow$ sites, respectively, the $h_{j,g}$
indices by $l_{j,g}$ indices, and the local rotated-electron double occupation operator
phase factors by one.

Before using properties 3-III, 6-III, and 7-III to confirm the validity of the local
pseudoparticle expressions (\ref{cnp})-(\ref{c3p}), let us check whether these
expressions conform to the requirement of property 5-III. According to such a
requirement, application of the off-diagonal generators of the $\eta$-spin $SU(2)$
algebra given in Eq. (7) of Ref. \cite{I} onto the superposition of the $2^{\nu}$
internal rotated-electron site distribution configurations of a $c\nu$ pseudoparticle
located in the rotated-electron lattice at position $[h_{j}\,a]$ must give zero. It is
straightforward to confirm from the analysis of the changes in the rotated-electron site
distribution configurations generated by these operators that this requirement is
fulfilled. The same is valid for application of the off-diagonal generators of the spin
$SU(2)$ algebra given in Eq. (8) of Ref. \cite{I} onto the superposition of the $2^{\nu}$
internal rotated-electron site distribution configurations of a $s\nu$ pseudoparticle
located in the rotated-electron lattice at position $[l_{j}\,a]$.

In order to confirm the validity of the pseudoparticle expressions
(\ref{cnp})-(\ref{c3p}), let us follow property 5-III and consider again a local charge
sequence with no Yang holons and constituted by a number $\nu$ of rotated-electron doubly
occupied sites and a number $\nu$ of rotated-electron unoccupied sites. If we request
that the rotated-electron distribution configurations of these $2\nu$ sites describe the
internal structure of a single $c\nu$ pseudoparticle, according to the restrictions
imposed by the Bethe-ansatz solution \cite{I}, the associated local charge sequence must
be characterized by the following numbers,

\begin{equation}
M_c =2\nu \, ; \hspace{0.5cm} N_{c\nu}=1 \, ; \hspace{0.5cm} N^h_{c\nu}=0 \, ;
\hspace{0.5cm} N_{c\nu'}=0 \, , \hspace{0.3cm} \nu'\neq\nu \, . \label{Ncn}
\end{equation}
Such a local charge sequence has a single $c\nu$ pseudoparticle and no $c\nu$
pseudoparticle {\it unoccupied sites}. This corresponds to an {\it effective $c\nu$
pseudoparticle lattice} constituted by a single {\it site} located in the
rotated-electron lattice at $[h_{1}\,a]$. That {\it site} is occupied by a $c\nu$
pseudoparticle. In this limiting case, in order to ensure periodic boundary conditions
for the original electronic problem it is required that the charge sequence must be
properly symmetrized by the method used in Ref. \cite{Geb}. Such a symmetrization
involves the operator ${\hat{\cal{T}}}_C$ of Eq. (\ref{Tcseq}).

The fulfillment of the requirement of property 5-III imposes that the superposition of
rotated-electron distribution configurations of the $\nu$ doubly occupied sites and $\nu$
unoccupied sites (and $\nu$ spin-down singly occupied sites and $\nu$ spin-up singly
occupied sites) of such a $c\nu$ pseudoparticle (and $s\nu$ pseudoparticle) must be
expressed in the form of a number $\nu$ of $h_{1,\,x} \leftrightarrow h_{1,\,x'}$ site
pairs (and $l_{1,\,x} \leftrightarrow l_{1,\,x'}$ site pairs) where $x$ and $x'$ are such
that $x'> x$ and have the values $x,x'=1,2,...,2\nu$. There is a number $(2\nu -1)!!$ of
possible different choices for these $\nu$ $h_{1,\,x} \leftrightarrow h_{1,\,x'}$ (and
$l_{1,\,x} \leftrightarrow l_{1,\,x'}$) site pairs. The use of requirement 6-III reveals
that only one of these $(2\nu -1)!!$ choices corresponds to the $c\nu$ pseudoparticle
(and $s\nu$ pseudoparticle). The remaining $(2\nu -1)!!-1$ choices can be expressed as a
superposition of several states described by different numbers $N_{c\nu'}$ of $c\nu'$
pseudoparticles such that $\sum_{\nu'=1}^{\nu}\nu'\,N_{c\nu'}=\nu$ (and $N_{s\nu'}$ of
$s\nu'$ pseudoparticles such that $\sum_{\nu'=1}^{\nu}\nu'\,N_{s\nu'}=\nu$). Moreover,
the fact that the symmetrized rotated-electron site distribution configuration which
describes the internal structure of the $c\nu$ pseudoparticle (and $s\nu$ pseudoparticle)
involves a number $\nu$ of site pairs implies that it corresponds to a state with charge
(and spin) momentum $k_C =\pi /a$ (and $k_S =\pi /a$). The possible values of such a
charge momentum are given in Eq. (\ref{kCkS}). For other values of charge momentum, such
a state cannot be described in terms of a number $\nu$ of site pairs and thus the
requirement introduced in property 5-III is not fulfilled.

For instance, when $\nu=1$ application of the basic properties 3-III, 5-III, and 6-III
leads uniquely to the following description of the $c1$ pseudoparticle,

\begin{equation}
\Bigl[\prod_{x=1}^{2}e^{i\pi
h_{1,\,x}{\hat{D}}_{1,\,x}}\Bigr]\,\sum_{\nu_C=0}^1\,e^{i\pi\nu_C}\,
\Bigl({\hat{\cal{T}}}_C\Bigr)^{\nu_C} \,(\bullet_{h_{1,\,1}},\,\circ_{h_{1,\,2}}) \, .
\label{c1SCS}
\end{equation}
This is the product of a rotated-electron double-occupation operator phase factor with a
symmetrized charge sequence. The charge sequence defined by Eq. (\ref{c1SCS}) can also be
written as,

\begin{equation}
\Bigl[\prod_{x=1}^{2}e^{i\pi
h_{1,\,x}{\hat{D}}_{1,\,x}}\Bigr]\,(1-{\hat{\cal{T}}}_{c1,\,1,\,1})\,
(\bullet_{h_{1,\,1}},\,\circ_{h_{1,\,2}}) \, , \label{c1alone}
\end{equation}
in agreement with the general expression (\ref{cnp}) and the local $c1$ pseudoparticle
expression (\ref{c1p}). For $j=1$, $x=1$, and $x'=2$ Fig. 1 describes the $h_{1,\,1}
\leftrightarrow h_{1,\,2}$ site pair associated with this $c1$ pseudoparticle.

Next, we consider the case $\nu=2$. The fulfillment of the requirement introduced in
property 5-III alone imposes in this case that the superposition of rotated-electron
distribution configurations of the two doubly occupied sites and two unoccupied sites
representative of a $c2$ pseudoparticle must be expressed in the form of two $h_{1,\,x}
\leftrightarrow h_{1,\,x'}$ site pairs where $x$ and $x'$ are such that $x'>x$ and have
the values $x,x'=1,2,3,4$. There are three possible different choices for the
superposition of rotated-electron distribution configurations formed by two site pairs.
These choices correspond to the following sets: (a) $\{h_{1,\,1} \leftrightarrow
h_{1,\,4}\, ;h_{1,\,2} \leftrightarrow h_{1,\,3}\}$, (b) $\{h_{1,\,1} \leftrightarrow
h_{1,\,3}\, ;h_{1,\,2} \leftrightarrow h_{1,\,4}\}$, and (c) $\{h_{1,\,1} \leftrightarrow
h_{1,\,2}\, ;h_{1,\,3} \leftrightarrow h_{1,\,4}\}$. According to property 7-III, the
charge sequence (c) is excluded, because it describes two $c1$ pseudoparticles and thus
there remain two possible choices for the $c2$ pseudoparticle. Figure 2 represents the
charge sequences corresponding to the choices (a) and (b). Each of the $h_{1,\,x}
\leftrightarrow h_{1,\,x'}$ site pairs are represented as in Fig. 1 with $j=1$. The
figure shows how the rotated-electron site distribution configurations of these two
charge sequences change as a result of a cyclic permutation which transforms the first
site of the charge sequence to the last site and multiplies the final state by a phase
factor of $-1$. This phase factor arises from the required $k_C =\pi /a$ charge momentum.
We note that the single $h_{1,\,1} \leftrightarrow h_{1,\,2}$ site pair descriptive of
the $c1$ pseudoparticle given in expressions (\ref{c1SCS}) and (\ref{c1alone}) transforms
into itself under such a transformation. On the other hand, in the present case the
rotated-electron site distribution configuration (a) transforms onto two $c1$
pseudoparticles whereas the rotated-electron site distribution configuration (b)
transforms onto itself. Thus, according to property 6-III only the rotated-electron site
distribution configuration (b) is properly symmetrized and describes the $c2$
pseudoparticle. Moreover, only this sequence can be written as a symmetrized charge
sequence of charge momentum $k_C =\pi /a$. Again, such a symmetrized charge sequence can
also be written in the form given in the general Eq. (\ref{cnp}) and in the local $c2$
pseudoparticle expression (\ref{c2p}). The two equivalent expressions read,

\begin{eqnarray}
& & \Bigl[\prod_{x=1}^{4}e^{i\pi
h_{1,\,x}{\hat{D}}_{j,\,x}}\Bigr]\,\sum_{\nu_C=0}^3\,e^{i\pi\nu_C}\,
\Bigl({\hat{\cal{T}}}_C\Bigr)^{\nu_C} (\bullet_{h_{1,\,1}},\,\bullet_{h_{1,\,2}},\,
\circ_{h_{1,\,3}},\,\circ_{h_{1,\,4}})\nonumber \\
& = & \Bigl[\prod_{x=1}^{4}e^{i\pi h_{1,\,x}{\hat{D}}_{j,\,x}}\Bigr]
\,\Bigl[\prod_{g=1}^{2}(1-{\hat{\cal{T}}}_{c2,\,1,\,g})\Bigr]\,
(\bullet_{h_{1,\,1}},\,\bullet_{h_{1,\,2}},\, \circ_{h_{1,\,3}},\,\circ_{h_{1,\,4}}) \, .
\label{c2alone}
\end{eqnarray}

\begin{figure*}
\includegraphics[width=7cm,height=5.00cm]{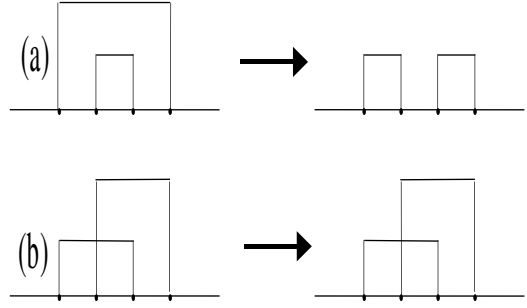}
\caption{\label{fig2} Graphical representation of charge sequences of a zero $\eta$-spin
energy eigenstate with no Yang holons and two rotated-electron doubly occupied sites and
two rotated-electron unoccupied sites only. The $h_{j,\,g} \leftrightarrow h_{j,\,g+\nu}$
site pairs are represented as in Fig. 1. We note that the relative height of different
site pairs has no physical meaning. The goal of using site pairs of different height is
just to clearly define the two sites of each pair. In figures (a) and (b) two possible
rotated-electron doubly occupied site/unoccupied  site pair distribution configurations
are represented. The figures show how these distribution configurations change as a
result of a cyclic permutation which transforms the first site of the charge sequence
onto its last site. The distribution configuration (b) transforms onto itself whereas the
distribution configuration (a) transforms onto two $c1$ pseudoparticles. Thus, according
to property 6-III the distribution configuration (b) describes a $c2$ pseudoparticle.
Note that the distribution configuration (a) describes two $c1$ pseudoparticles. The
distribution occupancies of the figure describe alternatively $s1$ and $s2$
pseudoparticles. In this case the spin sequences correspond to a zero spin energy
eigenstate with no HL spinons and with two spin-down singly occupied sites and two
spin-up singly occupied sites only.}
\end{figure*}

Let us now consider the case $\nu=3$. Again, the fulfillment of the requirement
introduced in property 5-III imposes that the superposition of rotated-electron
distribution configurations of the three doubly occupied sites and three unoccupied sites
representative of a $c3$ pseudoparticle should be expressed in the form of three
$h_{1,\,x} \leftrightarrow h_{1,\,x'}$ site pairs where $x$ and $x'$ are such that $x'>x$
and have the values $x,x'=1,2,3,4,5,6$. There are fifteen different possible choices for
superposition of rotated-electron site distribution configurations formed by three site
pairs. However, only one of these charge sequences can be written in the form of
symmetrized charge sequence of charge momentum $k_C =\pi /a$. For example, in Fig. 3 we
represent the charge sequences (a) $\{h_{1,\,1} \leftrightarrow h_{1,\,4}\, ;h_{1,\,2}
\leftrightarrow h_{1,\,5}\, ; h_{1,\,3} \leftrightarrow h_{1,\,6}\}$ and (b) $\{h_{1,\,1}
\leftrightarrow h_{1,\,3}\, ;h_{1,\,2} \leftrightarrow h_{1,\,5}\, ; h_{1,\,4}
\leftrightarrow h_{1,\,6}\}$. The figure shows how the rotated-electron site distribution
configurations of these two charge sequences change as a result of a cyclic permutation
which transforms the first site of the charge sequence to the last site and multiplies
the final state by a phase factor of $-1$. Both the rotated-electron site distribution
configuration (b) and all remaining possible rotated-electron site distribution
configurations constituted by three $h_{1,\,x} \leftrightarrow h_{1,\,x'}$ site pairs
except the rotated-electron site distribution configuration (a) do not transform onto
themselves. The latter rotated-electron site distribution configuration transforms onto
itself, as confirmed by analysis of the figure. Thus, according to property 6-III only
the rotated-electron site distribution configuration (a) is properly symmetrized and
describes the local $c3$ pseudoparticle. Only this sequence can be written in terms of a
symmetrized charge sequence of charge momentum $k_C =\pi /a$. Again, such symmetrized
charge sequence can also be written in the form given in the general Eq. (\ref{cnp}) and
in the local $c3$ expression (\ref{c3p}). The two equivalent expressions are the
following,

\begin{eqnarray}
\nonumber \\
& & \Bigl[\prod_{x=1}^{6}e^{i\pi
h_{1,\,x}{\hat{D}}_{j,\,x}}\Bigr]\,\{\sum_{\nu_C=0}^5\,e^{i\pi\nu_C}\,
\Bigl({\hat{\cal{T}}}_C\Bigr)^{\nu_c}\, (\bullet_{h_{1,\,1}},\,\bullet_{h_{1,\,2}},\,
\bullet_{h_{1,\,3}},\,\circ_{h_{1,\,4}},\,\circ_{h_{1,\,5}},\, \circ_{h_{1,\,6}})
\nonumber \\
& + & \sum_{\nu_C=0}^1\,e^{i\pi\nu_C}\,\Bigl({\hat{\cal{T}}}_C\Bigr)^{\nu_c}\,
(\circ_{h_{1,\,1}},\,\bullet_{h_{1,\,2}},\,
\circ_{h_{1,\,3}},\,\bullet_{h_{1,\,4}},\,\circ_{h_{1,\,5}},\,
\bullet_{h_{1,\,6}})\} \nonumber \\
& = & \Bigl[\prod_{x=1}^{6}e^{i\pi h_{1,\,x}{\hat{D}}_{j,\,x}}\Bigr]\,
\Bigl[\prod_{g=1}^{3}(1-{\hat{\cal{T}}}_{c3,\,1,\,g})\Bigr]\,
(\bullet_{h_{1,\,1}},\,\bullet_{h_{1,\,2}},\,
\bullet_{h_{1,\,3}},\,\circ_{h_{1,\,4}},\,\circ_{h_{1,\,5}},\, \circ_{h_{1,\,6}}) \, .
\label{c3alone}
\end{eqnarray}
Note that in this case one needs two symmetrized charge sequences to reach the $c3$
expression (\ref{c3p}) upon application of the operator ${\hat{\cal{T}}}_C$ onto these
charge sequences two and six times, respectively. However, in agreement with properties
5-III and 6-III, the state obtained from the superposition of these two charge sequences
is an eigenstate of the charge momentum operator with eigenvalue $k_C =\pi /a$. Also the
$c\nu$ pseudoparticles belonging to branches such that $\nu>3$ are a superposition of
several such symmetrized charge sequences. However, always the state obtained from the
superposition of several charge sequences is an eigenstate of the charge momentum
operator with eigenvalue $k_C =\pi /a$. Such a state can also be expressed in the general
form given in Eq. (\ref{cnp}).

\begin{figure*}
\includegraphics[width=7cm,height=5.00cm]{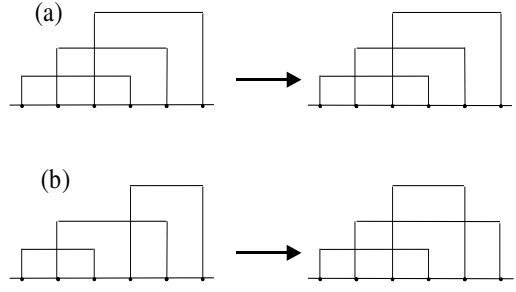}
\caption{\label{fig3} Graphical representation of charge sequences of a zero $\eta$-spin
energy eigenstate with no Yang holons and three rotated-electron doubly occupied sites
and three rotated-electron unoccupied sites only. The $h_{j,\,g} \leftrightarrow
h_{j,\,g+\nu}$ site pairs are represented as in Fig. 1. In figures (a) and (b) it is
shown how two possible rotated-electron doubly occupied site-unoccupied  site pair
distribution configurations change as a result of a cyclic permutation which transforms
the first site of the charge sequence onto its last site. The distribution configuration
(a) transforms onto itself whereas the distribution configuration (b) transforms onto a
non-equivalent distribution configuration. Thus, according to property 6-III the
distribution configuration (a) describes a $c3$ pseudoparticle. The figures represent
alternatively spin sequences of a zero spin energy eigenstate with no HL spinons and with
three spin-down singly occupied sites and three spin-up singly occupied sites only.}
\end{figure*}

For the case of charge (and spin) sequences constituted by several local $c\nu$
pseudoparticles (and local $s\nu$ pseudoparticles) belonging to one or several
$\nu=1,2,...$ branches and by Yang holons (and HL spinons) the internal structure of the
local $c\nu$ pseudoparticles (and local $s\nu$ pseudoparticles) is given by Eq.
(\ref{cnp}) (and Eq. (\ref{snp})). This result follows from the basic property 7-III. We
will in the ensuing section confirm that, in agreement with property 8-III, this ensures
the periodic boundary conditions of the original electronic problem provided that the
bare-momentum energy eigenstates are suitable Fourier-transform superpositions of the
local charge sequences with bare-momentum values obeying Eqs. (\ref{pbccp}) and
(\ref{pbcanp}).

However, whether inside the domain of $2\nu$ sites of a local $c\nu$ pseudoparticle there
can be sites with index $j_h$ such that $j_{h_{j,\,1}}<j_h<j_{h_{j,\,2\nu}}$ and which
sites are occupied by Yang holons or other local $c\nu'$ pseudoparticles remains an open
question. This problem is absent in the limiting case we considered here where the charge
sequence is constituted by a number $2\nu$ of sites of the rotated-eletron lattice. This
issue is closely related to the definition of the local $c\nu$ pseudoparticle
charge-sequence {\it unoccupied sites}, which is discussed in the following. A similar
problem occurs for the spin sequences.

\subsection{THE LOCAL $\alpha\nu$ PSEUDOPARTICLE {\it UNOCCUPIED SITES} AND THE
MOVING PSEUDOPARTICLE ELEMENTARY STEPS}

In the definition of the $N^h_{c\nu}$ charge-sequence {\it unoccupied sites} of a local
$c\nu$ pseudoparticle we start by considering charge sequences where $N_{c\nu}=1$ and all
remaining local $c\nu'$ pseudoparticles except the specific local $c\nu$ pseudoparticle
keep their charge-sequence positions ${\bar{h}}_j$ defined in Eq. (\ref{bstringC}). Below
we call {\it steady} local $c\nu'$ pseudoparticles the former local pseudoparticles whose
charge-sequence positions ${\bar{h}}_j$ remain unchanged. This concept does not apply for
charge sequences where $N_{c\nu}>1$, when the specific local $c\nu$ pseudoparticle passes
a pseudoparticle belonging to the same branch. Indeed, in this case the two objects are
indiscernible. However, also for such charge sequences the number of unoccupied sites
remains well defined, as discussed below. Obviously, any local pseudoparticle can be
chosen to be the {\it moving} pseudoparticle. This just refers to a set of different
charge sequences where the charge-sequence position of that quantum object changes
whereas the charge-sequence positions of the remaining local pseudoparticles remain
unchanged. We note that according to Eq. (\ref{bstringC}), the charge-sequence position
of a local $c\nu$ pseudoparticle can remain unchanged in spite of the {\it movements} of
some of its $2\nu$ sites, as we confirm below. Concerning the definition of the
$N^h_{c\nu}$ charge-sequence {\it unoccupied sites} of a local $c\nu$ pseudoparticle,
from the use of properties 1-III to 8-III and of other properties of the model we find
the following distribution configuration properties:\vspace{0.5cm}

1-IV Let us consider a set of rotated-electron doubly occupied and/or rotated-electron
unoccupied sites located at positions $[j_h\,a]$ where according to Eqs. (\ref{hl}) and
(\ref{chcl}) the index $h$ has the values $h=1,2,...,[N_a-N_c]$. Moreover, we consider
that all of these sites are located inside the domain of sites of a local $c\nu$
pseudoparticle and thus their indices $j_h$ are such that
$j_{h_{j,\,1}}<j_h<j_{h_{j,\,2\nu}}$ and $j_h\neq j_{h_{j,\,x}}$ where $x=1,2,...,2\nu$.
This property states that the set of rotated-electron doubly occupied/unoccupied sites of
index $j_h$ obeying these inequalities cannot correspond to sites occupied by Yang holons
or to any of the $2\nu'$ sites of the charge sequence associated with local $c\nu'$
pseudoparticles belonging to branches such that $\nu'\geq\nu$. Thus, such a set of
rotated-electron doubly occupied or unoccupied sites can only correspond to sub-sets of
$2\nu'$ sites of the charge sequence associated with local $c\nu'$ pseudoparticles
belonging to branches such that $\nu'<\nu$.\vspace{0.5cm}

2-IV Let us consider a local $c\nu$ pseudoparticle and a local $c\nu'$ pseudoparticle
belonging to the same charge sequence and to branches such that $\nu<\nu'$. Moreover, we
consider that there are no charge-sequence local pseudoparticles inside the $2\nu'$-site
domain of the $c\nu'$ pseudoparticle other than the $c\nu$ pseudoparticle. (The
description of cases involving more than two local pseudoparticles can be easily reached
by generalization of the two-pseudoparticle situation considered here.) Then the $2\nu$
sites of the charge sequence associated with the local $c\nu$ pseudoparticle are located
either outside the domain of $2\nu'$ sites of the local $c\nu'$ pseudoparticle limited in
the left and right ends by the sites of index $h_{j,\,1}$ and $ h_{j,\,2\nu'}$,
respectively, or are all located inside such a domain. If the $2\nu$ sites are located
inside the domain of these $2\nu'$ sites there is a number $2\nu'-1$ of permitted
positions for the $c\nu$ pseudoparticle. In each of these permitted positions all of the
$2\nu$ sites of the local $c\nu$ pseudoparticle are located between two first neighboring
sites of indices $h_{j,x}$ and $h_{j',x+1}$ of the local $c\nu'$ pseudoparticle. These
permitted positions correspond to the neighboring sites of indices $h_{j,x}$ and
$h_{j',x+1}$ such that $x=1,...,2\nu'-1$.\vspace{0.5cm}

3-IV Let us consider again the local $c\nu$ pseudoparticle and the local $c\nu'$
pseudoparticle referred to in property 2-IV. The charge-sequence position ${\bar{h}}_j$
given in Eq. (\ref{bstringC}) of such a steady local $c\nu'$ pseudoparticle remains
unchanged in all possible $2\nu'-1$ positions of the local $c\nu$ pseudoparticle while
passing its $2\nu'$-site domain. The inverse statement is even stronger. It is required
that when a local $c\nu'$ pseudoparticle belonging to a branch such that $\nu'>\nu$
passes a steady local $c\nu$ pseudoparticle, the position in the rotated-electron lattice
of all the $2\nu$ sites of the latter pseudoparticle must remain unchanged. Moreover, the
$2\nu'-1$ possible different positions of the $2\nu$ sites of the local $c\nu$
pseudoparticle inside the site domain of the local $c\nu'$ pseudoparticle must be the
same when the local $c\nu$ pseudoparticle passes the steady local $c\nu'$ pseudoparticle
from its left to its right-hand side and vice versa. Also the relative positions of the
two quantum objects when the local $c\nu$ pseudoparticle passes the steady local $c\nu'$
pseudoparticle and when the local $c\nu'$ pseudoparticle passes the steady local $c\nu$
pseudoparticle must be the same. However, note that although these relative positions are
the same, when a local $c\nu'$ pseudoparticle belonging to a branch such that $\nu'>\nu$
passes a steady local $c\nu$ pseudoparticle it does not occupy any of the $2\nu$ sites of
the rotated-electron lattice occupied by the latter local pseudoparticle. Thus, the
positions in the lattice of these $2\nu$ sites remain indeed unchanged. A last
requirement is that the positions of all $2\nu'$ sites of a steady local $c\nu'$
pseudoparticle belonging to a branch such that $\nu'>\nu$ must be the same in the
rotated-electron site distribution configurations before and after it is passed by the
moving local $c\nu$ pseudoparticle. If the local $c\nu$ pseudoparticle is moving from the
left to the right-hand side of the steady local $c\nu'$ pseudoparticle, {\it before} and
{\it after} is meant here as the rotated-electron site distribution configurations where
the $2\nu$ sites of the former quantum object are located outside the domain of $2\nu'$
sites of the steady local pseudoparticle and on its right and left hand sides,
respectively.\vspace{0.5cm}

4-IV Corresponding rules are also valid for the case of the spin sequences and local
$s\nu$ pseudoparticles.\vspace{0.5cm}

A simple example of a domain constituted by thirteen sites of a permitted charge sequence
is represented in Fig. 4. In such a figure the $h_{j,\,g} \leftrightarrow h_{j,\,g+\nu}$
site pairs of the local $c\nu$ pseudoparticles are represented as in Figs. 1 to 3. The
six sites with no vertical lines correspond to unoccupied sites associated with Yang
holons. There are two local $c1$ pseudoparticles and a local $c2$ pseudoparticle in the
charge-sequence domain represented in the figure.

\begin{figure*}
\includegraphics[width=7cm,height=5.00cm]{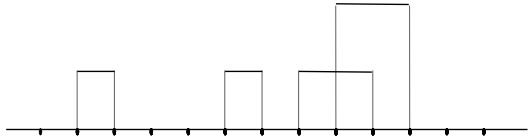}
\caption{\label{fig4} Graphical representation of a domain of a charge sequence of the
rotated-electron lattice including thirteen sites. The $h_{j,\,g} \leftrightarrow
h_{j,\,g+\nu}$ site pairs are represented as in Fig. 1. The six sites with no vertical
lines correspond to Yang holons. There are two local $c1$ pseudoparticles described by
$h_{j,\,1} \leftrightarrow h_{j,\,2}$ site pairs and a local $c2$ pseudoparticle
described by a $h_{j,\,1} \leftrightarrow h_{j,\,3}$ and a $h_{j,\,2} \leftrightarrow
h_{j,\,4}$ site pairs. Alternatively, if we replace the Yang holons by HL spinons and the
$c\nu$ pseudoparticles by $s\nu$ pseudoparticles the figure represents the domain of a
spin sequence with six HL spinons, two $s1$ pseudoparticles, and a $s2$ pseudoparticle.}
\end{figure*}

In order to confirm the validity of the counting of rotated-electron site distribution
configurations leading to expression (\ref{NanNS}), let us use the properties 1-IV to
4-IV in the classification of the different possible positions of a local $c\nu$
pseudoparticle in the rotated-electron lattice. Our analysis can be easily generalized to
the case of a $s\nu$ pseudoparticle. Our study includes the introduction of the concept
of {\it $2\nu$-leg caterpillar step}, where the $2\nu$-leg caterpillar is nothing but the
$2\nu$-site pseudoparticle block associated with the moving $c\nu$ pseudoparticle. For
simplicity, we consider a charge sequence such that $N_{c\nu}=1$ for the $\nu$ branch.
The case of sequences with $N_{c\nu}>1$ is considered below. As mentioned above, the
possible positions of such a pseudoparticle correspond to a number $N^*_{c\nu}$ of
different charge sequences where the charge-sequence positions of all remaining quantum
objects remain unchanged. Thus, these $N^*_{c\nu}$ charge sequences differ in the
position of the $c\nu$ pseudoparticle relative to the other quantum objects. The above
property 3-IV states that the charge-sequence position ${\bar{h}}_j$ given in Eq.
(\ref{bstringC}) of all quantum objects except the position of the {\it moving} $c\nu$
pseudoparticle must remain unchanged in all these $N^*_{c\nu}$ charge sequences. That is
obviously true for steady local $c\nu'$ pseudoparticles belonging to branches such that
$\nu'<\nu$. Let us confirm that the charge-sequence position of a steady local $c\nu'$
pseudoparticle belonging to a branch such that $\nu'>\nu$ indeed remains unchanged when
it is passed by a local $c\nu$ pseudoparticle, in spite of changes of the positions in
the rotated-electron lattice of some of its $2\nu'$ sites.

We start by considering the case of a local $c{\bar{\nu}}$ pseudoparticle where
$\nu={\bar{\nu}}$ denotes the largest $\nu$ value of the $c\nu$ pseudoparticle branches
with finite occupancy in the charge sequence. According to Eq. (B.7) of Ref. \cite{I}, in
this case the number of local $c{\bar{\nu}}$ pseudoparticle {\it unoccupied sites} is
given by $N^h_{c{\bar{\nu}}} = L_{c}$ and equals the number of Yang holons. The different
possible positions of a $c{\bar{\nu}}$ pseudoparticle can be achieved by elementary steps
in the charge sequence where each of its $2{\bar{\nu}}$ sites moves forward by a single
site of the charge sequence. If the movement is from the left to the right-hand side, the
$2{\bar{\nu}}\,{\rm th}$ pseudoparticle site of index $h_{j,\,2{\bar{\nu}}}$ moves into
the site previously occupied by a Yang holon. The remaining $2{\bar{\nu}}-1$
pseudoparticle sites of index $h_{j,\,x}$ move into the site previously associated with
the pseudoparticle site of index $h_{j,\,x+1}$ where $x=1,...,2{\bar{\nu}}-1$. Finally,
the Yang holon removed from the site newly associated with the pseudoparticle site of
index $h_{j,\,2{\bar{\nu}}}$ moves into the site previously associated with the first
pseudoparticle site of index $h_{j,\,1}$. We call such an elementary collective step of
the $2{\bar{\nu}}$ pseudoparticle sites {\it $2{\bar{\nu}}$-leg caterpillar step}.
According to such an analogy, the $2{\bar{\nu}}$ {\it legs} refer to these sites and the
{\it caterpillar} refers to the local pseudoparticle. While each of these legs moves
forward by a charge-sequence site, the whole caterpillar itself also moves forward by a
single charge-sequence site. The net result of such a $2{\bar{\nu}}$-leg caterpillar step
is that the compact domain of $2{\bar{\nu}}$ pseudoparticle sites interchanges position
with a Yang holon. When such a moving $c{\bar{\nu}}$ pseudoparticle passes a steady local
$c\nu$ pseudoparticle, each of its $2{\bar{\nu}}$ sites jump the $2\nu$ sites of the
latter local pseudoparticle. Thus, the $2{\bar{\nu}}$-leg caterpillar moves forward like
the $2\nu$ sites of such a steady local $c\nu$ pseudoparticle do not exist in the charge
sequence. As we discuss below, the construction of the corresponding effective
$c{\bar{\nu}}$ pseudoparticle lattice mentioned in property 8-III involves omission of
the sites of the rotated-electron lattice belonging to the $2\nu$-site domains of such a
local $c\nu$ pseudoparticles. Moreover, for such an effective $c{\bar{\nu}}$
pseudoparticle lattice the $2{\bar{\nu}}$-site domain of the $c{\bar{\nu}}$
pseudoparticles is {\it seen} as a point-like occupied site. Thus, in the particular case
of the ${\bar{\nu}}$ branch such an effective lattice has
$N^*_{c{\bar{\nu}}}=L_c+N_{c{\bar{\nu}}}$ sites and a number $N^h_{c{\bar{\nu}}}=L_c$
free of $c{\bar{\nu}}$ pseudoparticles, as we confirm below.

Let us consider now the case of a general local $c\nu$ pseudoparticle belonging to a
branch such that $1\leq\nu<{\bar{\nu}}$. For simplicity, we keep the assumption that
$N_{c\nu}=1$ for the $\nu$ branch. We consider a given charge sequence where the $c\nu$
pseudoparticle is located in the rotated-electron lattice at position $[h_j\,a]$. It
follows from property 4-III that the $N^h_{c\nu}$ remaining possible positions of the
local $c\nu$ pseudoparticle in the charge sequence define the positions of the local
$c\nu$ pseudoparticle charge-sequence {\it unoccupied sites}. Out of these $N^h_{c\nu} =
L_{c} + 2\sum_{\nu'=\nu +1}^{\infty} (\nu' -\nu) N_{c\nu'}$ possible positions, $L_c$ is
the number of positions associated with Yang holons. Again, the different positions of
the moving local $c\nu$ pseudoparticle can be achieved by $2\nu$-leg caterpillar steps.
The net result of such a $2\nu$-leg caterpillar step is that the compact domain
constituted by the $2\nu$ $c\nu$ pseudoparticle sites interchanges position with a Yang
holon or with a site of a $h_{j',\,g'}\leftrightarrow h_{j',\,g'+\nu'}$ pair of a steady
local $c\nu'$ pseudoparticle belonging to a branch such that $\nu'>\nu$. A $2\nu$-leg
caterpillar step is defined here as for the above moving $c{\bar{\nu}}$ pseudoparticle.
If the movement is from the left to the right-hand side, the local $c\nu$ pseudoparticle
site of index $h_{j,\,2\nu}$ moves into the site previously occupied by a Yang holon or
into one of the two sites previously associated with a local $c\nu'$ pseudoparticle
$h_{j',\,g'} \leftrightarrow h_{j',\,g'+\nu'}$ site pair. The remaining $2\nu-1$ sites of
the local $c\nu$ pseudoparticle of index $h_{j,\,x}$ such that $x=1,...,2\nu-1$, move
into the site previously associated with the local $c\nu$ pseudoparticle site of index
$h_{j,\,x+1}$ where again $x=1,...,2\nu-1$. Finally, the Yang holon or $c\nu'$
pseudoparticle site associated with the local $h_{j',\,g'} \leftrightarrow
h_{j',\,g'+\nu'}$ site pair, which was removed from the site newly associated with the
local $c\nu$ pseudoparticle site of index $h_{1,\,2\nu}$, moves into the site previously
associated with the local $c\nu$ pseudoparticle site of index $h_{j,\,1}$. Again, the net
result of such a $2\nu$-leg caterpillar step is that the local $c\nu$ pseudoparticle
moves forward by a single charge-sequence site.

Let us consider that in the initial rotated-electron site distribution configuration the
moving $c\nu$ pseudoparticle is located on the left-hand side of the steady local $c\nu'$
pseudoparticle. Thus, in the final state just after passing the latter pseudoparticle,
the $c\nu$ pseudoparticle will occupy in the charge sequence a domain of $2\nu$ sites
located on the right-hand side of the $c\nu'$ pseudoparticle site domain. According to
property 3-IV, in both the initial and final rotated-electron site distribution
configurations all $2\nu'$ sites of the latter quantum object remain the same. For
simplicity, we consider that in the initial rotated-electron site distribution
configuration, the $2\nu$ charge-sequence sites located just on the right-hand side of
the steady local $c\nu'$ pseudoparticle are occupied by Yang holons. However, our results
can be generalized to rotated-electron site distribution configurations where these sites
are occupied by a third pseudoparticle. Another case which we could consider is when the
moving local pseudoparticle passes two steady local $c\nu'$ and $c\nu''$ pseudoparticles
such that one of them is located within the site domain of the other one. From the
systematic use of the properties 1-IV to 4-IV one can describe the movements of a local
$c\nu$ pseudoparticle when it passes any rotated-electron site distribution configuration
involving several steady local pseudoparticles.

According to property 2-IV, the $2\nu$ sites of a local $c\nu$ pseudoparticle can have
$2\nu'-1$ different positions inside the $2\nu'$-site domain of a local $c\nu'$
pseudoparticle belonging to a branch such that $\nu'>\nu$. We consider that in the
initial rotated-electron site distribution configuration, the local $c\nu$ and $c\nu'$
pseudoparticles are located side by side in the charge sequence. Thus, these two quantum
objects occupy together a compact $2(\nu+\nu')$-site domain. By simple counting arguments
it is straightforward to realize that in order to reach each of these permitted
positions, each of the $2\nu$ sites of the local $c\nu$ pseudoparticle must jump $2\nu$
sites of the charge sequence, somewhere inside the $2\nu'$-site domain of the steady
local $c\nu'$ pseudoparticle. This result is consistent with the expression of the number
$2\sum_{\nu'=\nu +1}^{\infty} (\nu' -\nu) N_{\alpha\nu'}$ of the available positions for
the moving local $c\nu$ pseudoparticle involving sites of steady local $c\nu'$
pseudoparticles belonging to branches such that $\nu'>\nu$. This expression reveals that
when located inside the site domain of such a pseudoparticle, each of the $2\nu$ sites of
the local $c\nu$ pseudoparticle can move into $2(\nu' -\nu)$ sites out of the $2\nu'$
total number of sites of the steady local $c\nu'$ pseudoparticle. This agrees with the
above result obtained from counting arguments alone. We note that the number of
charge-sequence sites jumped by each local $c\nu$ caterpillar leg equals precisely
$2\nu$, the number of sites of such a $c\nu$ pseudoparticle. Thus, these jumps ensure
that the position in the rotated-electron lattice of all the $2\nu'$ sites of the steady
local $c\nu'$ pseudoparticle remains unchanged after the $c\nu$ passes it, as requested
by the property 3-IV. This result also applies to the general situation when a moving
local $c\nu$ pseudoparticle passes a domain containing several local steady $c\nu'$
pseudoparticles belonging to branches such that $\nu'>\nu$. Also in the general situation
that each of the legs of the moving caterpillar uses $2(\nu' -\nu)$ possible sites of
each steady local $c\nu'$ pseudoparticle, out of a total of $2\nu'$ sites.

The question which arises from the occurrence of the above jump mechanism is which are
the $2\nu$ sites of the steady local $c\nu'$ pseudoparticle that are jumped by each site
of the moving local $c\nu$ pseudoparticle. In the following analysis of the site
distribution configurations of the $c\nu$ pseudoparticle when it passes the steady local
$c\nu'$ pseudoparticle we provide explicit information about the location of its $2\nu$
sites. The corresponding location of the $2\nu'$ sites of the steady local $c\nu'$
pseudoparticle is implicity given if one recalls that in all occupancy configurations
reported below both pseudoparticles occupy a compact charge-sequence domain of
$2(\nu+\nu')$ sites. While these occupancy configurations involve movements of the
$2\nu'$ sites of the $c\nu'$ pseudoparticle, according to the property 3-IV the steady
character of that object follows from the invariance of its charge-sequence position
${\bar{h}}_j$, as given by Eq. (\ref{bstringC}), which remains the same in all such
occupancy configurations. The requirements mentioned in that property provide the needed
information for the definition of the $2\nu'-1$ permitted positions of the local $c\nu$
pseudoparticle inside the steady local $c\nu'$ pseudoparticle: To reach these permitted
positions, each site of the local $c\nu$ pseudoparticle must perform two independent
jumps of $\nu$ sites in the charge sequence, as described below. The possible positions
of a local $c\nu$ pseudoparticle when it passes from the left to the right-hand side of a
steady local $c\nu'$ pseudoparticle are the following: First the local $c\nu$
pseudoparticle performs a $2\nu$-leg caterpillar step. The net result of such an
elementary collective step is that the compact domain of $2\nu$ sites of the local $c\nu$
pseudoparticle interchanges position with the site of index $h_{j',\,1}$ of the local
$c\nu'$ pseudoparticle. In the reached rotated-electron site distribution configuration
all the $2\nu$ sites of the local $c\nu$ pseudoparticle are located between the sites of
index $h_{j',\,1}$ and $h_{j',\,2}$ of the steady local $c\nu'$ pseudoparticle. This
procedure is repeated $\nu'-1$ times until the $2\nu$ sites of the $c\nu$ pseudoparticle
are located between the sites of index $h_{j',\,\nu'-1}$ and $h_{j',\,\nu'}$ of the
steady local $c\nu'$ pseudoparticle. In each of the corresponding $2\nu$-leg caterpillar
steps the local $c\nu$ pseudoparticle moves forward by a single lattice site of the
charge sequence. However, in the $\nu'\,{\rm th}$ step each site of the $c\nu$
pseudoparticle jumps $\nu$ sites of the charge sequence. This jump brings it to a
rotated-electron site distribution configuration where the charge-sequence position
${\bar{h}}_j$ of the local $c\nu$ pseudoparticle given in Eq. (\ref{bstringC}) coincides
with that of the steady local $c\nu'$ pseudoparticle. In order to reach the next
permitted position each of the $2\nu$ local $c\nu$ pseudoparticle sites jump again a
number $\nu$ of lattice sites of the charge sequence. This brings the local $c\nu$
pseudoparticle to a distribution configuration where its $2\nu$ sites are located between
the sites of index $h_{j',\,1+\nu'}$ and $h_{j',\,2+\nu'}$ of the steady local $c\nu'$
pseudoparticle. These two jumps are followed by $\nu'-1$ $2\nu$-leg caterpillar steps
where the $c\nu$ pseudoparticle moves again by a single lattice site of the charge
sequence. The net result of each of these elementary collective steps is that the compact
domain of $2\nu$ $c\nu$ pseudoparticle sites interchanges position with one of the sites
of the $c\nu'$ pseudoparticle. Finally, the last step brings the local $c\nu$
pseudoparticle to a distribution configuration where it is located outside and on the
right-hand side of the domain of $2\nu'$ sites of the steady local $c\nu'$
pseudoparticle. The same intermediate positions are reached when the $c\nu$
pseudoparticle passes the domain of $2\nu'$ sites of the $c\nu'$ pseudoparticle from the
right to the left-hand side. Such events are illustrated in Fig. 5 for the case when a
local $c1$ pseudoparticle passes a steady local $c2$ pseudoparticle from the left to the
right-hand side. If instead we consider that the local $c1$ pseudoparticle is steady,
Fig. 5 illustrates the possible positions of a local $c2$ pseudoparticle when it passes
from the right to the left-hand side a steady local $c1$ pseudoparticle. In this case the
position in the effective electronic lattice of the two sites of the $c1$ pseudoparticle
remains unchanged and the different relative positions of the two quantum objects are
reached by movements of the $c2$ pseudoparticle. Importantly, we emphasize that in this
case the $c2$ pseudoparticle does not occupy the sites occupied by the $c1$
pseudoparticle. Again this is consistent with property 3-IV.

\begin{figure*}
\includegraphics[width=7cm,height=5.00cm]{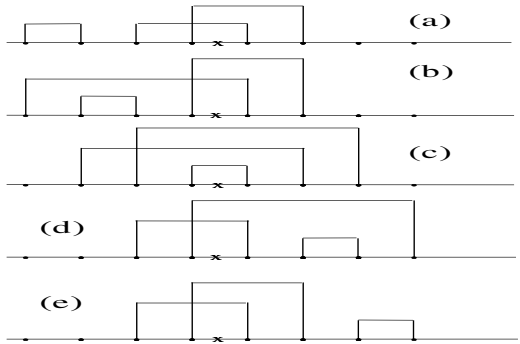}
\caption{\label{fig5} An illustration of the possible positions of a local $c1$
pseudoparticle when it passes from the left to the right-hand side of a steady local $c2$
pseudoparticle. The $h_{j,\,g} \leftrightarrow h_{j,\,g+\nu}$ site pairs are represented
as in Fig. 1. In the five rotated-electron site distribution configurations (a)-(e) the
charge-sequence position of the steady $c2$ pseudoparticle is labeled by a $X$ and
remains unchanged. If instead we consider a steady local $c1$ pseudoparticle, figures
(a)-(e) illustrate the possible positions of a local $c2$ pseudoparticle when it passes
from the right to the left-hand side of a steady local $c1$ pseudoparticle. However, in
this second case the charge-sequence position of the local $c1$ pseudoparticle should
remain unchanged and the $X$ point of figures (a)-(e) should be moved accordingly. In
both cases these five distribution configurations correspond to five different charge
sequences. Alternatively, the figure represents the corresponding electron site
distribution configurations of a $s1$ pseudoparticle and a $s2$ pseudoparticle.}
\end{figure*}

We have just studied a situation where a local $c\nu$ pseudoparticle passes a steady
local $c\nu'$ pseudoparticle belonging to a branch such that $\nu'>\nu$. For simplicity,
we assumed that in the initial rotated-electron site distribution configuration the
domain of $2\nu$ charge-sequence sites located just on the right-hand side of the steady
local $c\nu'$ pseudoparticle is occupied by Yang holons. However, our results can be
generalized to more complicated situations. An example is a situation where a local
$c\nu$ pseudoparticle passes a steady local $c\nu'$ pseudoparticle which has as first
neighbor in the charge sequence a second steady local $c\nu''$ pseudoparticle. Another
example is when a local $c\nu$ pseudoparticle passes a steady local $c\nu'$
pseudoparticle containing inside its $2\nu'$-site domain a third steady local $c\nu''$
pseudoparticle belonging to a branch such that $\nu''<\nu'$. These more general
situations can be described by the use of the properties 1-IV to 4-IV. One always finds
that in order to move forward, the $2\nu$ sites of the local $c\nu$ pseudoparticle use
$2(\nu' -\nu)$ sites out of the total number of $2\nu'$ sites of each steady local
$c\nu'$ pseudoparticle belonging to branches such that $\nu'>\nu$. Moreover, when passing
the central region of the $2\nu'$-site domain of such a local $c\nu'$ pseudoparticle, the
$2\nu$ legs of the $c\nu$ caterpillar always perform the two $\nu$-site jumps described
above.

Finally, we consider the general case of a charge sequence such that $N_{\alpha\nu}>1$
for the $\alpha\nu$ branch of the moving local pseudoparticle. In this case such an
object passes local pseudoparticles belonging to the same branch. We consider that in the
initial rotated-electron site distribution configuration, the moving $c\nu$
pseudoparticle and a second $c\nu$ pseudoparticle are located side by side in the charge
sequence. Thus, these two quantum objects occupy together a compact $4\nu$-site domain.
Let us suppose that the local $c\nu$ pseudoparticle is moving from the left to the
right-hand side. It follows from the indiscernible character of the two objects that in
this case they move side by side, both performing $2\nu$ $2\nu$-leg caterpillar steps.
This brings such an objects to a distribution configuration where the $c\nu$
pseudoparticle on the left-hand side has the same position than that on the right-hand
side had in the initial distribution configuration. Since the two quantum objects are
indiscernible, the moving $c\nu$ pseudoparticle is now that located on the right-hand
side. By counting the involved site distribution configurations one realizes that they
are consistent with the number of unoccupied sites being $N^h_{c\nu} = L_{c} +
2\sum_{\nu'=\nu +1}^{\infty} (\nu' -\nu) N_{c\nu'}$.

The charge sequence (and the spin sequence) is obtained by omission of the
rotated-electron singly occupied sites (and rotated-electron doubly occupied/unoccupied
sites). Such a procedure follows from the independent conservation of the charge and spin
sequence. We note that each of the $c\nu$ pseudoparticle (and $s\nu$ pseudoparticle)
branch occupancy configurations are also independently conserved. Thus, it follows from
our above analysis that the construction of the effective $c\nu$ pseudoparticle lattice
mentioned in property 8-III also involves omission of the sites of the rotated-electron
lattice which are jumped by the movements of a $c\nu$ pseudoparticle, since these sites
are not {\it seen} by such a moving quantum object. The omitted sites correspond to the
$2\nu'$-site domains of local $c\nu'$ pseudoparticles of branches such that $\nu'<\nu$
and to $2\nu$-site domains belonging to the $2\nu'$ sites of each $c\nu'$ pseudoparticle
such that $\nu'>\nu$. Moreover, for such an effective $c\nu$ pseudoparticle lattice the
$2\nu$-site domain of the $c\nu$ pseudoparticles is {\it seen} as a point-like occupied
site. Thus, such an effective lattice has
$N^*_{c\nu}=L_c+\sum_{\nu'>\nu}2(\nu'-\nu)\,N_{c\nu'}+N_{c\nu}$ sites, a number
$N^h_{c\nu}=L_c+2\sum_{\nu'>\nu}(\nu'-\nu)\,N_{c\nu'}$ of which are unoccupied  of $c\nu$
pseudoparticles, as we confirm below. A similar analysis holds for the $s\nu$
pseudoparticles.

\subsection{COMPLETE SET OF CHARGE, SPIN, AND $c$ PSEUDOPARTICLE
SEQUENCE LOCAL STATES}

We have just found the rotated-electron distribution configurations of the $2\nu$ sites
which describe a local $\alpha\nu$ pseudoparticle. Moreover, we considered the
$N^h_{\alpha\nu}$ charge-sequence {\it unoccupied sites} of such a local $\alpha\nu$
pseudoparticle. Let us consider again the case of the local charge sequences and
associated local $c\nu$ pseudoparticles. A result which follows from our findings is that
indeed the positions of the $N^h_{c\nu}$ charge-sequence {\it unoccupied sites} of each
local $c\nu$ pseudoparticle branch are fully determined by the position of the $c$
pseudoparticles and $c\nu$ pseudoparticles alone. Our results thus confirm that for fixed
values of the numbers $N_{c\nu}$ and $N^*_{c\nu}$ the number of occupancy configurations
of the local $c\nu$ pseudoparticles is given by Eq. (\ref{NanNS}) for $\alpha =c$. It
follows that we can classify the local $c\nu$ pseudoparticle occupancy configurations for
$\nu=1,2,...$ branches with finite occupancy in a given state by providing the indices
$h_j$ given in Eqs. (\ref{stringC}). These indices correspond to the $N_{c\nu}$
pseudoparticle positions provided in Eq. (\ref{localcnp}). Our results also confirm that
the number of local pseudoparticle occupancy configurations of a CPHS ensemble subspace
is indeed given by Eq. (\ref{Ncphs}). Finally, since our results can be generalized to
spin sequences as well, they also validate the use of the representation associated with
the local state defined in Eqs. (\ref{localst})-(\ref{args}). Such a local state
represents each of the different local pseudoparticle occupancy configurations whose
number for a specific CPHS ensemble subspace is given in Eq. (\ref{Ncphs}). Our results
also confirm that the set of these local states whose number is given in Eq.
(\ref{Ncphs}) constitutes a complete set of states in the corresponding CPHS ensemble
subspace.

Thus, for CPHS ensemble subspaces with no $-1/2$ Yang holons and no $-1/2$ HL spinons the
local states defined in Eqs. (\ref{localst})-(\ref{args}) constitute a complete set of
states. Our main goal here is the generalization of these local states to CPHS ensemble
subspaces with finite occupancies of both $\pm 1/2$ Yang holons and $\pm 1/2$ HL spinons.
We start by expressing the complete set of local states defined in Eqs.
(\ref{localst})-(\ref{args}) in terms of the following more basic states,

\begin{equation}
\vert(j_{1},\,j_2,...,j_{N_c});\, \{(\bullet_{h_{j,\,1}},...,\bullet_{h_{j,\,\nu}},
\,\circ_{h_{j,\,1+\nu}},..., \circ_{h_{j,\,2\nu}})\};\,\{(\downarrow_{l_{j,\,1}},...,
\downarrow_{l_{j,\,\nu}},\,\uparrow_{l_{j,\,1+\nu}},...,
\uparrow_{l_{j,\,2\nu}})\}\rangle \, , \label{initial}
\end{equation}
where the charge sequence refers to the following sets of rotated-electron distribution
configurations of doubly-occupied and unoccupied sites,

\begin{eqnarray}
& & \{(\bullet_{h_{j,\,1}},...,\bullet_{h_{j,\,\nu}}, \,\circ_{h_{j,\,1+\nu}},...,
\circ_{h_{j,\,2\nu}})\} =
(\bullet_{h_{1,\,1}},\,\circ_{h_{1,\,2}}),\,(\bullet_{h_{2,\,1}},\,\circ_{h_{2,\,2}}),...,
(\bullet_{h_{N_{c1}},\,1},\,\circ_{h_{N_{c1},\,2}});
\nonumber \\
& & (\bullet_{h_{1,\,1}},\,\bullet_{h_{1,\,2}},\,
\circ_{h_{1,\,3}},\,\circ_{h_{1,\,4}}),\,(\bullet_{h_{2,\,1}},\,\bullet_{h_{2,\,2}},\,
\circ_{h_{2,\,3}},\,\circ_{h_{2,\,4}})
,\,...,(\bullet_{h_{N_{c2},\,1}},\,\bullet_{h_{N_{c2},\,2}},\,
\circ_{h_{N_{c2},\,3}},\,\circ_{h_{N_{c2},\,4}});
...;\nonumber \\
& & (\bullet_{h_{1,\,1}},...,\bullet_{h_{1,\,\nu}}, \,\circ_{h_{1,\,1+\nu}},...,
\circ_{h_{1,\,2\nu}}),\, (\bullet_{h_{2,\,1}},...,\bullet_{h_{2,\,\nu}},
\,\circ_{2_{j,\,1+\nu}},..., \circ_{h_{2,\,2\nu}}),...,\nonumber \\
& & (\bullet_{h_{N_{c\nu},\,1}},...,\bullet_{h_{N_{c\nu},\,\nu}},\,
\circ_{h_{N_{c\nu},\,\nu+1}},...,\circ_{h_{N_{c\nu},\,2\nu}}) ;\,... \label{argcn}
\end{eqnarray}
and the same follows for the spin sequence with the index $h$ replaced by the index $l$,
rotated-electron doubly occupied sites $\bullet$ replaced by spin-down rotated-electron
singly occupied sites $\downarrow$, and sites free of rotated electrons $\circ$ replaced
by spin-up rotated-electron singly occupied sites $\uparrow$. The states defined in Eqs.
(\ref{initial})-(\ref{argcn}) provide the position of the $2\nu$ sites associated with
each of the $\alpha\nu$ pseudoparticles whose branches have finite occupancy in the
corresponding charge and spin sequence. However, we note that these states include only
rotated-electron site distribution configurations of the simple type given in Eqs.
(\ref{lcnp}) and (\ref{lsnp}). The general charge and spin sequences can be obtained from
these states by application of suitable operators, as we discussed above. These are thus
the most basic and simple states. Our first task is expressing the states defined in Eqs.
(\ref{localst})-(\ref{args}) in terms of these simple states.

Each local $c\nu$ and $s\nu$ pseudoparticle is described by the superposition of
$2^{\nu}$ rotated-electron site distribution configurations given in Eqs. (\ref{cnp}) and
(\ref{snp}), respectively. The type of configuration superposition given in these
equations is common to the description of all local $\alpha\nu$ pseudoparticles of a
charge or spin sequence. Thus, the proper description of the charge and spin sequence of
each of the states defined in Eqs. (\ref{localst})-(\ref{args}) involves the
superposition of $2^{\textstyle\sum_{\alpha
=c,s}\sum_{\nu=1}^{\infty}\nu\,N_{\alpha\nu}}$ such a rotated-electron site distribution
configurations. It follows that these normalized local states can be expressed as
follows,

\begin{eqnarray}
& & \vert(j_{1},\,j_2,...,j_{N_c});\,
\{(h_{1},\,h_2,...,h_{N_{c\nu}})\};\,\{(l_{1},\,l_2,...,l_{N_{s\nu}})\}\rangle
\nonumber \\
& = & \Bigl\{2^{-\textstyle[\sum_{\alpha
=c,s}\sum_{\nu=1}^{\infty}\nu\,N_{\alpha\nu}]/2}\Bigr\}\,
\Bigl[\prod_{\nu'=1}^{\infty}\prod_{j'=1}^{N_{c\nu'}}\prod_{x=1}^{2\nu'}e^{i\pi
h_{j',\,x}{\hat{D}}_{j',\,x}}\Bigr]\,\Bigl[\prod_{\alpha
=c,s}\prod_{\nu''=1}^{\infty}\prod_{j''=1}^{N_{\alpha\nu''}}
\prod_{g''=1}^{\nu''}(1-{\hat{\cal{T}}}_{\alpha\nu'',\,j'',\,g''})\Bigr]\nonumber
\\
& \times & \vert(j_{1},\,j_2,...,j_{N_c});\,
\{(\bullet_{h_{j,\,1}},...,\bullet_{h_{j,\,\nu}}, \,\circ_{h_{j,\,1+\nu}},...,
\circ_{h_{j,\,2\nu}})\};\,\{(\downarrow_{l_{j,\,1}},...,
\downarrow_{l_{j,\,\nu}},\,\uparrow_{l_{j,\,1+\nu}},...,
\uparrow_{l_{j,\,2\nu}})\}\rangle \, , \label{prima}
\end{eqnarray}
where the operator ${\hat{D}}_{j',\,x}$ measures the number of rotated-electron doubly
occupied sites in the charge-sequence site of index $h_{j',\,x}$ (whose value is given by
$1$ or $0$) and the transformation laws for application of the operator
${\hat{\cal{T}}}_{\alpha\nu'',\,j'',\,g''}$ were given above and are illustrated in Eqs.
(\ref{Tc}) and (\ref{Ts}).

The local states given in both Eqs. (\ref{localst})-(\ref{args}) and Eq. (\ref{prima})
provide a complete basis of states for CPHS ensemble subspaces spanned by lowest-weight
states of both the $\eta$-spin and spin $SU(2)$ algebras. Let us now consider the general
case of states associated with CPHS ensemble subspaces spanned by states with both finite
occupancies of $\pm 1/2$ Yang holons and $\pm 1/2$ HL spinons. The description of the
local charge and spin sequences with finite occupancies of $\pm 1/2$ Yang holons and $\pm
1/2$ HL spinons, respectively, is quite similar to the ones we studied above. For
simplicity, let us consider first the case of the charge sequences. According to property
1-III the $-1/2$ Yang holons and $+1/2$ Yang holons correspond to single rotated-electron
doubly occupied lattice sites and single rotated-electron unoccupied  lattice sites of
such a charge sequence, respectively. For a lowest weight state (and highest weight
state) of the $\eta$-spin algebra all Yang holons have the same $\eta$-spin projection
$+1/2$ (and $-1/2$) and correspond to rotated-electron unoccupied sites (and
rotated-electron doubly occupied sites). The application of the off-diagonal generators
of the $\eta$-spin $SU(2)$ algebra given in Eq. (7) of Ref. \cite{I} leads to flips of
the Yang holons $\eta$-spins \cite{I}. Following property 5-III, in order to achieve the
Yang holon occupancy configurations required by the $\eta$-spin $SU(2)$ algebra, we find
that each local charge sequence is described by a superposition of $C_c$ Yang holon
occupancy configurations where,

\begin{equation}
C_{\alpha} = {L_{\alpha}\choose L_{\alpha,\,-1/2}} = {L_{\alpha}!\over
L_{\alpha,\,-1/2}!\,L_{\alpha,\,+1/2}!} \, ; \hspace{0.5cm} \alpha = c,\,s \, .
\label{NLc}
\end{equation}
Here $L_{c}=2S_c$ (and $L_{s}=2S_s$) stands for the total number of Yang holons (and HL
spinons) in the charge (and spin) sequence. All the ${L_c\choose L_{c,\,-1/2}}$ states
associated with these configurations have the same set of $L_c$ lattice sites occupied by
Yang holons and the same numbers of $-1/2$ Yang holons and $+1/2$ Yang holons. However,
these states differ in the distributions of the $-1/2$ Yang holons and $+1/2$ Yang holons
over the $L_c$ lattice sites. The same analysis holds for spin sequences with finite
occupancies of $\pm 1/2$ HL spinons. Thus, there are $\prod_{\alpha =c,s}(L_{\alpha}+1)$
states with precisely the same occupancy configurations of local $c$ pseudoparticles and
local $\alpha\nu$ pseudoparticles. Each of these occupancy configurations is described by
a state of form given in Eq. (\ref{prima}). On the other hand, we denote each of the
$\prod_{\alpha =c,s}(L_{\alpha}+1)$ general states by,

\begin{equation}
\vert(L_{c,-1/2},\,L_{s,-1/2});\,(j_{1},\,j_2,...,j_{N_c});\,
\{(h_{1},\,h_2,...,h_{N_{c\nu}})\};\,\{(l_{1},\,l_2,...,l_{N_{s\nu}})\}\rangle \, ,
\label{nlwss}
\end{equation}
where $L_{c,-1/2}$ and $L_{s,-1/2}$ are the corresponding numbers of $-1/2$ Yang holons
and $-1/2$ HL spinons, respectively. We note that Eq. (\ref{LcsLWS}) remains valid for
the general local states (\ref{nlwss}). Thus, the values of the numbers $L_c$ and $L_s$
are determined by the values of the numbers of $c$ pseudoparticles and of $\alpha\nu$
pseudoparticles. The values of the numbers $L_{c,-1/2}$ and $L_{s,-1/2}$ were added to
the local states (\ref{nlwss}) because they are needed for the specification of the
corresponding CPHS ensemble subspace. In contrast, the values of the numbers $L_{c,+1/2}$
and $L_{s,+1/2}$ are dependent and given by $L_{c,+1/2}=L_c-L_{c,-1/2}$ and
$L_{s,+1/2}=L_s-L_{c,-1/2}$ and are not explicitly provided in the expression of the
local states (\ref{nlwss}). Each of these normalized states is given by,

\begin{eqnarray}
& & \vert (L_{c,-1/2},\,L_{s,-1/2});\,(j_{1},\,j_2,...,j_{N_c});\,
\{(h_{1},\,h_2,...,h_{N_{c\nu}})\};\,\{(l_{1},\,l_2,...,l_{N_{s\nu}})\}\rangle
\nonumber \\
& = & 2^{-\textstyle[\sum_{\alpha
=c,s}\sum_{\nu=1}^{\infty}\nu\,N_{\alpha\nu}]/2}\nonumber \\
& \times & \Bigl[\prod_{\alpha
=c,s}{\Bigl({\hat{S}}_{\alpha}^{\dag}\Bigr)^{L_{\alpha,\,-1/2}}\over
\sqrt{C_{\alpha}}}\Bigr]\,
\Bigl[\prod_{\nu'=1}^{\infty}\prod_{j'=1}^{N_{c\nu'}}\prod_{x=1}^{2\nu'}e^{i\pi
h_{j',\,x}{\hat{D}}_{j',\,x}}\Bigr]\,\Bigl[\prod_{\alpha''
=c,s}\prod_{\nu''=1}^{\infty}\prod_{j''=1}^{N_{\alpha''\nu''}}
\prod_{g''=1}^{\nu''}(1-{\hat{\cal{T}}}_{\alpha''\nu'',\,j'',\,g''})\Bigr]
\nonumber \\
& \times & \vert(j_{1},\,j_2,...,j_{N_c});\,
\{(\bullet_{h_{j,\,1}},...,\bullet_{h_{j,\,\nu}}, \,\circ_{h_{j,\,1+\nu}},...,
\circ_{h_{j,\,2\nu}})\};\,\{(\downarrow_{l_{j,\,1}},...,
\downarrow_{l_{j,\,\nu}},\,\uparrow_{l_{j,\,1+\nu}},...,
\uparrow_{l_{j,\,2\nu}})\}\rangle \, ,
\label{completes}
\end{eqnarray}
where the operators ${\hat{S}}_{\alpha}^{\dag}$ with $\alpha =c,s$ are the off-diagonal
generators defined in Eqs. (7) and (8) of Ref. \cite{I} and the corresponding
normalization constants $C_{\alpha}$ are given in Eq. (\ref{NLc}). Application of these
off-diagonal generators onto the local states given in Eq. (\ref{prima}) leaves the
rotated-electron site distribution configurations of the $\alpha\nu$ pseudoparticles
invariant and generates $\eta$-spin and spin flips in the $+1/2$ Yang holons and $+1/2$
HL spinons, respectively, as requested by property 5-III.

The set of all local states of general form given in Eq. (\ref{completes}) provides a
complete basis of states in any CPHS ensemble subspace of the Hilbert space. However,
these states do not ensure the periodic boundary conditions of the original electronic
problem. Such conditions are ensured by the construction of the energy eigenstates as
Fourier-transform superpositions of local states of general form given in Eq.
(\ref{completes}), as discussed in the ensuing section.

\section{SEPARATION OF THE $\alpha\nu$ PSEUDOPARTICLE TRANSLATIONAL -
INTERNAL DEGREES OF FREEDOM: EFFECTIVE $\alpha\nu$ PSEUDOPARTICLE LATTICES AND THE ENERGY
EIGENSTATES}

In this section we finally reach a precise definition for the concept an effective
$\alpha\nu$ pseudoparticle lattice. The precise definition of such a concept is a
necessary condition for the construction of the energy eigenstates as Fourier-transform
superpositions of the local states introduced in the previous section. The concept of an
effective $\alpha\nu$ pseudoparticle lattice is related to the separation of the
translational and internal degrees of freedom of these quantum objects, as we explain
below. As mentioned in property 8-III, in addition to the effective $c$ pseudoparticle
lattice, there is an effective $\alpha\nu$ pseudoparticle lattice for each $\alpha\nu$
pseudoparticle branch. The spatial coordinate associated with these lattices is the
conjugate of the pseudoparticle bare-momentum $q_j$, which is a good quantum number of
the many-electron problem. As discussed in Appendix B of Ref. \cite{I}, this quantum
number is associated with the Bethe-ansatz solution. The Fourier transforms which relate
the local pseudoparticles introduced in this paper to the bare-momentum pseudoparticles
obtained from analysis of the Bethe-ansatz solution in Ref. \cite{I} involve the spatial
coordinate of the effective pseudoparticle lattices.

Besides the introduction of the effective pseudoparticle lattices, in this section we
also provide explicit expressions for the energy eigenstates in terms of Slater
determinants which involve Fourier-transform superpositions of the local states
introduced in the previous section. This clarifies the relation of the energy eigenstates
obtained from the use of the Bethe-ansatz solution \cite{Lieb,Takahashi} and $\eta$-spin
and spin $SU(2)$ symmetries \cite{HL,Yang89} to the rotated-electron site distribution
configurations that emerge from the electron - rotated-electron unitary transformation
\cite{I}.

\subsection{TRANSLATIONAL - INTERNAL DEGREES OF FREEDOM SEPARATION
AND THE EFFECTIVE $\alpha\nu$ PSEUDOPARTICLE LATTICES}

For simplicity, let us again concentrate mainly on the case of the $c\nu$
pseudoparticles. The separation of the $\alpha\nu$ pseudoparticle translational and
internal degrees of freedom can be defined in terms of the following general properties
which are obtained from the above properties 1-III to 8-III and 1-IV to 3-IV, and other
features of the model:\vspace{0.5cm}

1-V One can separate the $2\nu$-site internal structure of a local $c\nu$ pseudoparticle
from the problem of the description of the movements of these quantum objects in the
rotated-electron lattice. Such separation leads to the concept of effective
pseudoparticle lattice, whose spatial coordinates are associated with the pseudoparticle
translational degrees of freedom only. The motion of a $c\nu$ pseudoparticle involves
$2\nu$-leg caterpillar steps, whereas the $2\nu$-site domains of these objects play the
role of the occupied {\it sites} of an effective $c\nu$ pseudoparticle lattice. From the
point of view of the motion of these objects in such an effective $c\nu$ pseudoparticle
lattice, these $2\nu$-site domains are point-like sites without internal structure.
Roughly speaking, these point-like pseudoparticle occupied sites correspond to the {\it
center of mass} of the $2\nu$-site block associated with each local $c\nu$
pseudoparticle. This description corresponds to a separation of the pseudoparticle
translational and internal degrees of freedom. The latter degrees of freedom are
described by the $2\nu$-site rotated-electron distribution configurations studied in the
previous section. That internal structure is such that the $c\nu$ pseudoparticle
transforms as a $\eta$-spin zero quantum object under application of the off-diagonal
generators of the $\eta$-spin $SU(2)$ algebra. In addition, such an internal structure is
a necessary condition for the fulfilment of the periodic boundary conditions of the
original electronic problem. On the other hand, the $c\nu$ pseudoparticle translational
degrees of freedom are closely related to the spatial coordinates of the effective
pseudoparticle lattice. These spatial coordinates play the role of conjugate variable
relative to the bare momentum $q_j$. Such a bare momentum obeys Eq. (\ref{pbcanp}) to
ensure periodic boundary conditions for the original electronic problem. An important
point is that the pseudoparticle spatial coordinate occupancy configuration of each
$c\nu$ pseudoparticle branch with finite occupancy in a given energy eigenstate is
independently conserved. As the charge (and spin) sequence corresponds to the
rotated-electron doubly occupied and unoccupied sites (and rotated-electron singly
occupied sites) only, one can also introduce an independent effective $c\nu$
pseudoparticle lattice for each of these occupied pseudoparticle branches, whose
coordinates correspond to some of the sites of the local charge sequence only.
\vspace{0.5cm}

2-V Within the above separation of the pseudoparticle translational and internal degrees
of freedom, the moving $c\nu$ pseudoparticle {\it sees} the $2\nu$-site domains of each
of the local pseudoparticles belonging to the same branch as point-like occupied
pseudoparticle sites. On the other hand, according to the property 3-IV a moving $c\nu$
pseudoparticle jumps the $2\nu'$-site domains representative of each local $c\nu'$
pseudoparticles belonging to branches such that $\nu'<\nu$. This property can be
understood as follows: Since in its movements the $c\nu$ pseudoparticle {\it sees} the
$2\nu$-site domains of the remaining $c\nu$ pseudoparticles of the same branch as
point-like unreachable occupied pseudoparticle sites, such a $2\nu$-site domain width is
the smallest width {\it seen} by the $c\nu$ pseudoparticle. Thus, such a width plays the
role of a site domain {\it width uncertainty}. As a result, pseudoparticle site domains
of width smaller than the $2\nu$-site domain are not {\it seen} by the $c\nu$
pseudoparticle. This is consistent with the exact property that the moving $c\nu$
pseudoparticle does not {\it see} ({\it i.e.} jumps) smaller $2\nu'$-site $c\nu'$
pseudoparticle domains corresponding to branches such that $\nu'<\nu$. On the other hand,
again according to property 3-IV, the $2\nu$-leg caterpillar step movements of the $c\nu$
pseudoparticle uses $2(\nu'-\nu)$ sites only, out of the $2\nu'$-site domain of each
local $c\nu'$ pseudoparticle belonging to branches such that $\nu'>\nu$. Thus, again
consistently with the above site domain {\it width uncertainty}, in their motion
throughout the rotated-electron lattice, a local $c\nu$ pseudoparticle does not {\it see}
a number $2\nu$ of sites out of the $2\nu'$ sites of such a $\nu'>\nu$ local $c\nu'$
pseudoparticle.\vspace{0.5cm}

3-V The concepts of an unoccupied  site and a lattice constant of the effective $c\nu$
pseudoparticle lattice follow from the above analysis. For the moving $c\nu$
pseudoparticle all the $N^h_{c\nu}$ charge-sequence {\it unoccupied sites} corresponding
to a number $L_c$ of $\pm 1/2$ Yang holons and a number
$2\sum_{\nu'=\nu+1}^{\infty}(\nu'-\nu)\,N_{c\nu'}$ of sites belonging to the $2\nu'$-site
domains of local $c\nu'$ pseudoparticles of the charge sequence such that $\nu'>\nu$ are
seen as equivalent and indiscernible point-like unoccupied sites of a {\it effective
$c\nu$ pseudoparticle lattice}. Moreover, according to the property 1-V such a quantum
object also {\it sees} the $N_{c\nu}$ $2\nu$-site domains of pseudoparticles belonging to
the $c\nu$ pseudoparticle branch as equivalent and indiscernible point-like occupied
sites. Thus, while running through all its possible positions in the rotated-electron
lattice, the moving $c\nu$ pseudoparticle {\it sees} all the
$N^*_{c\nu}=N_{c\nu}+N^h_{c\nu}$ sites of the {\it effective $c\nu$ pseudoparticle
lattice} as equivalent and indiscernible point-like sites. Therefore, these sites are
equally spaced for the effective $c\nu$ pseudoparticle lattice. The value of the
corresponding lattice constant is uniquely defined as follows: In its movements a $c\nu$
pseudoparticle {\it does not see} (i) the $2\sum_{\nu'=1}^{\nu}\nu'\,N_{c\nu'}$ sites
occupied by $c\nu'$ pseudoparticles belonging to branches such that $\nu'\leq\nu$; (ii) a
number $2\nu\,\sum_{\nu'=\nu+1}^{\infty}N_{c\nu'}$ of sites belonging to the $2\nu'$-site
domains of $c\nu'$ pseudoparticles belonging to branches such that $\nu'>\nu$; and (iii)
the $N_c$ sites singly occupied by rotated electrons. Thus, the $c\nu$ pseudoparticle
{\it jumps} all the above sites of the rotated-electron lattice and {\it sees} the
$2\nu$-site pseudoparticle domains of the $c\nu$ pseudoparticle branch as point-like
occupied sites. On the other hand, in order to pass all the
$N^*_{c\nu}=N_{c\nu}+N^h_{c\nu}$ sites of the effective $c\nu$ pseudoparticle lattice and
return to its original position, the $c\nu$ pseudoparticle should run through a distance
which equals the length $L$ of the rotated-electron lattice. Therefore, a necessary
condition to ensure the periodic boundary conditions of the original electronic problem
is that the length of the effective $c\nu$ pseudoparticle lattice must equal the length
$L$ of the rotated-electron lattice. This determines uniquely the value of the lattice
constant of the effective $c\nu$ pseudoparticle lattice which reads
$a_{c\nu}=L/N^*_{c\nu}$. That the length of the effective $c\nu$ pseudoparticle lattice
is $L$ is consistent with the spacing $q_{j+1}-q_j=2\pi/L$ given in Eq. (B.2) of Ref.
\cite{I} for the corresponding pseudoparticle discrete bare-momentum values provided by
the exact Bethe-ansatz solution. Such a discrete bare momentum $q_j$ is the conjugate of
the coordinate $x_j=a_{c\nu}\,j$ of the effective $c\nu$ pseudoparticle lattice where
$j=1,2,...,N^*_{c\nu}$. \vspace{0.5cm}

4-V A similar analysis is valid for the local $s\nu$ pseudoparticles, provided that the
above mentioned $N_c$ sites singly occupied by rotated electrons are replaced by the
$[N_a-N_c]$ sites doubly occupied by rotated electrons and free of rotated electrons. The
lattice constant of the effective $s\nu$ pseudoparticle lattice is given by
$a_{s\nu}=L/N^*_{s\nu}$. On the other hand, the effective $c$ pseudoparticle lattice has
$N_a$ lattice sites and its lattice constant $a$ is the same as for the rotated-electron
lattice. The positions of the $N_c$ $c$ pseudoparticles and $N^h_c=[N_a-N_c]$ $c$
pseudoparticle holes in this lattice equal the corresponding positions of the
rotated-electron singly occupied sites and rotated-electron doubly occupied/unoccupied
sites, as already discussed in previous sections.\vspace{0.5cm}

5-V The set of pseudoparticle occupancy configurations of the effective $c$
pseudoparticle, $c\nu$ pseudoparticle, and $s\nu$ pseudoparticle lattices belonging to
branches $\nu=1,2,...$ with finite occupancy in a given local state of form
(\ref{completes}) together with the numbers $L_{c,\,-1/2}$ and $L_{s,\,-1/2}$ of such a
state contain the same information as the corresponding description of the same local
state in terms of the rotated-electron site distribution configurations.\vspace{0.5cm}

Since the value of the number $N^*_{\alpha\nu}$ defined by Eqs. (B.6), (B.7), and (B.11)
of Ref. \cite{I} is distinct for different CPHS ensemble subspaces, it follows from the
expression of the effective $\alpha\nu$ pseudoparticle lattice constants,

\begin{equation}
a_{\alpha\nu} = a\,{N_a\over N^*_{\alpha\nu}} =  {L\over N^*_{\alpha\nu}} \, ,
\label{aan}
\end{equation}
that the value of such constants changes accordingly.

Let us introduce the following notation for the spatial coordinates of the effective $c$
pseudoparticle lattice,

\begin{equation}
x_j =a\,j \, , \hspace{0.5cm} j=1,2,3,...,N_a \, , \label{xc}
\end{equation}
where the index $j$ was called $j_l$ and $j_h$ in Eq. (\ref{chcl}) for the $c$
pseudoparticle occupied sites ($l=1,2,3,...,N_c$) and unoccupied sites
($h=1,2,3,...,[N_a-N_c]$), respectively. Moreover, let us introduce the following
notation for the spatial coordinates of the effective $\alpha\nu$ pseudoparticle
lattices,

\begin{equation}
x_j =a_{\alpha\nu}\, j \, , \hspace{0.5cm} j=1,2,3,...,N^*_{\alpha\nu} \, . \label{xeg}
\end{equation}
The bare-momentum limiting values given in Eq. (B.14) of Ref. \cite{I} for the
$\alpha\nu$ pseudoparticle bands can be expressed in terms of the corresponding lattice
constants $a_{\alpha\nu}$ as follows,

\begin{equation}
q_{\alpha\nu} = {\pi\over a_{\alpha\nu}}[1-1/N_a] \approx {\pi\over a_{\alpha\nu}} \, .
\label{qagaan}
\end{equation}
Also the $c$ pseudoparticle limiting bare-momentum values given in Eqs. (B.16) and (B.17)
of the same paper can be written as,

\begin{equation}
q_c^{\pm} \approx \pm q_c \, ; \hspace{0.5cm} q_c = {\pi\over a} \, . \label{qceac}
\end{equation}
Thus, the domain of available pseudoparticle bare-momentum values corresponds to an
effective {\it first-Brillouin zone} associated with an underlying effective
pseudoparticle lattice.

In reference \cite{I} it is found that both the electronic lattice and the momentum
operator (\ref{Popel}), which is the generator of the spatial translations in such a
lattice, remain invariant under the electron - rotated-electron unitary transformation.
This justifies why the electronic and rotated-electron lattices are identical. Therefore,
the lattice spatial coordinates occupied by the rotated electrons correspond to
real-space coordinates. Since we have just shown that the effective pseudoparticle
lattices refer directly to the rotated-electron lattice, also the space coordinates of
the former lattices refer to real-space coordinates. The pseudoparticle bare momentum is
the conjugate of the spatial coordinate of the corresponding effective pseudoparticle
lattice. Thus, it follows from the above invariance and relations that the pseudoparticle
bare-momentum refers to real momentum. This is confirmed by the form of Eq. (36) of Ref.
\cite{I} for the momentum in terms of the pseudoparticle bare-momentum occupancy
configurations. Indeed, the expression of such an equation is additive in the bare
momentum for all pseudoparticle branches.

The ground state plays an important role in the study of one- and two-electron spectral
functions \cite{V}. Thus, let us use the ground-state Eqs. (B.11), (C.12)-(C.14), (C.24),
and (C.25) of Ref. \cite{I} in order to find the corresponding values for the effective
$\alpha\nu$ pseudoparticle lattice constants in the case of electronic densities $n$ and
spin densities $m$ such that $0< n\,a< 1$ and $0< m\,a< n\,a$, respectively. We find that
in the case of such a state these constants read,

\begin{equation}
a_{c\nu}^0 = {1\over \delta} \, ; \hspace{1cm} a_{s1}^0 = {1\over n_{\uparrow}} \, ;
\hspace{1cm} a_{s\nu}^0 = {1\over m} \, , \label{acanGS}
\end{equation}
where $\delta =[1-na]/a$ is the doping concentration away from half filling. Moreover,
the ground-state number $N^*_{\alpha\nu}$ given in Eqs. (B.6), (B.7), and (B.11) of Ref.
\cite{I} and the ground-state limiting bare-momentum values given in Eq. (C.12)-(C.14) of
the same paper can be written in terms of the effective pseudoparticle lattice constants
as follows,

\begin{equation}
N^*_{\alpha\nu}={L\over a_{\alpha\nu}^0} \, ; \hspace{0.5cm} q^0_{\alpha\nu} = {\pi\over
a_{\alpha\nu}^0} \, .\label{qcanGSefa}
\end{equation}
We note that when the effective pseudoparticle lattice constants given in Eq.
(\ref{acanGS}) diverge, as for $a_{c\nu}^0$ as $\delta a=[1-na]\rightarrow 0$ and
$a_{s\nu}^0$ as $ma=[n_{\uparrow}-n_{\downarrow}]a\rightarrow 0$ when $\nu >1$, the value
of the corresponding number $N^*_{\alpha\nu}$ given in Eq. (\ref{qcanGSefa}) is zero.
This just means that in these limits the corresponding effective pseudoparticle lattices
do not exist for states belonging to the ground-state CPHS ensemble subspace.

The relation of the effective pseudoparticle lattices introduced here to previous results
on the model in the limit $U/t\rightarrow\infty$ is briefly discussed in the Appendix.

\subsection{CONSTRUCTION OF THE ENERGY EIGENSTATES}

In order to relate the description of the energy eigenstates in terms of pseudoparticle
occupancy configurations obtained from the use of the Bethe-ansatz solution and
$\eta$-spin and spin symmetries to the representation of the same states in terms of
rotated-electron site distribution configurations, there is a last goal to be reached.
That is the construction of the energy eigenstates in terms of the local charge, spin,
and $c$ pseudoparticle sequences studied in previous sections. In section IV we expressed
the local states given in both Eqs. (\ref{localst})-(\ref{args}) and Eq. (\ref{prima}) in
terms of rotated-electron site distribution configurations and showed that they provide a
complete basis of states. In order to relate these local states to the energy eigenstates
we should first express the former states in terms of local $c$ and $\alpha\nu$
pseudoparticle occupancy configurations in the corresponding effective pseudoparticle
lattices. We then Fourier transform the obtained states into bare-momentum space $q_j$
with respect to the spatial coordinates of the effective $c$ and $\alpha\nu$
pseudoparticle lattices.

As in the case of Eq. (\ref{chcl}) for the effective $c$ pseudoparticle lattice, we
denote by $j_l$ and $j_h$ where $l=1,2,...,N_{\alpha\nu}$ and
$h=1,2,...,N^h_{\alpha\nu}$, respectively, the occupied and unoccupied sites,
respectively, of the effective $\alpha\nu$ pseudoparticle lattices. The spatial
coordinate of these occupied and unoccupied pseudoparticle sites is given by,

\begin{equation}
x_{j_{l}} = a_{\alpha\nu}\,j_l \, , \hspace{0.25cm} l=1,2,...,N_{\alpha\nu} \, ;
\hspace{1cm} x_{j_{h}} = a_{\alpha\nu}\,j_h \, , \hspace{0.25cm}
h=1,2,...,N^h_{\alpha\nu} \, , \label{aeloe}
\end{equation}
where $j_l$ is the index which defines the position of the occupied pseudoparticle sites
and $j_h$ the position of the unoccupied pseudoparticle sites. We note that the
unoccupied sites of the effective pseudoparticle lattices correspond to the sites left
over by the occupied sites and vice versa. Thus, we can uniquely specify a given
effective pseudoparticle lattice site distribution configuration by providing the
position of the occupied sites (or unoccupied sites) only. Here we choose the
representation in terms of the positions in the effective $c$ and $\alpha\nu$
pseudoparticle lattices of the occupied sites $x_{j_{l}}$ of Eqs. (\ref{chcl}) and
(\ref{aeloe}). In such a representation the states given in Eq. (\ref{completes}) are
denoted as,

\begin{equation}
\vert(L_{c,-1/2},\,L_{s,-1/2});\,(x_{j_{1}},\,x_{j_2},...,x_{j_{N_c}});\,
\{(x_{j_{1}},\,x_{j_2},...,x_{j_{N_{c\nu}}})\};\,\{(x_{j_{1}},\,x_{j_2},...,x_{j_{N_{s\nu}}})\}\rangle
\, . \label{ellss}
\end{equation}
Here $(x_{j_{1}},\,x_{j_2},...,x_{j_{N_c}})$ is the set of spatial coordinates associated
with the set of indices $(j_{1},\,j_2,...,j_{N_c})$ given in Eq. (\ref{lcp}). These
spatial coordinates correspond to the set of sites occupied by $c$ pseudoparticles in the
corresponding effective lattice. Moreover,

\begin{equation}
\{(x_{j_{1}},\,x_{j_2},...,x_{j_{N_{\alpha\nu}}})\} =
(x_{j_1},\,x_{j_2},...,x_{j_{N_{\alpha 1}}});\, (x_{j_{1}},\,x_{j_2},...,x_{j_{N_{\alpha
2}}});\,(x_{j_{1}},\,x_{j_2},...,x_{j_{N_{\alpha 3}}});... \, , \label{eanel}
\end{equation}
gives the spatial coordinates of each $\alpha\nu$ pseudoparticle effective lattice with
finite occupancy.

We emphasize that from analysis of the specific rotated-electron site distribution
configurations relative to a given state of form (\ref{prima}) one can construct by use
of the above properties 1-V to 3-V the $c$ pseudoparticle, $c\nu$ pseudoparticle, and
$s\nu$ pseudoparticle occupancy configurations in the effective $c$, $c\nu$, and $s\nu$
pseudoparticle lattices, respectively, of the corresponding state of form (\ref{ellss}).
This relation was already mentioned in the above property 5-V and the inverse statement
is obviously also true. Thus, Eqs. (\ref{prima}) and (\ref{ellss}) refer to two different
representations of the same local states. These local states constitute a complete set of
states. However, the form of these local states does not ensure the periodic boundary
conditions of the original electronic problem.

The representation of the local states in terms of the effective pseudoparticle lattice
occupancy configurations given in Eq. (\ref{ellss}) is the most suitable starting point
for construction of the energy and momentum eigenstates. Such a construction is fulfilled
by Fourier transforming the local states given in Eq. (\ref{ellss}) into bare-momentum
space with respect to the spatial coordinates of the effective pseudoparticle lattices
given in Eqs. (\ref{xc}) and (\ref{xeg}). Provided that the discrete bare-momentum values
obey the boundary conditions associated with Eqs. (\ref{pbccp}) and (\ref{pbcanp}), the
form of the obtained states ensures the periodic boundary conditions of the original
electronic problem. Such procedures lead to the following expression for the energy
eigenstates in terms of the local states of Eq. (\ref{ellss}),

\begin{eqnarray}
& & \vert (L_{c,-1/2},\,L_{s,-1/2});\,(q_{j_1},\,q_{j_2},...,q_{j_{N_c}});\,
\{(q_{j_1},\,q_{j_2},...,q_{j_{N_{c\nu}}})\};\,\{(q_{j_1},\,q_{j_2},...,q_{j_{N_{s\nu}}})\}\rangle
\nonumber \\
& = & N_a^{-N_c/2}\Bigl(\prod_{\alpha' =c,s}\prod_{\nu'=1}^{\infty}
[N^*_{\alpha'\nu'}]^{-N_{\alpha'\nu'}/2}\Bigr)\,
\sum_{j_{1}<j_2<...<j_{N_c}}\sum_{{\cal{P}}}\,(-1)^{\cal{P}} \,e^{\textstyle
(ia\sum_{l=1}^{N_c}j_{{\cal{P}}(l)}\,q_{j_{l}})}\nonumber
\\
& \times &
\Bigl[\prod_{\nu'=1}^{\infty}\,\Bigl(\sum_{j_{1}<j_2<...<j_{N_{c\nu'}}}\sum_{{\cal{P}}}\,(-1)^{\cal{P}}
\,e^{\textstyle (ia_{c\nu}\sum_{l=1}^{N_{c\nu'}}j_{{\cal{P}}(l)}\,[{\pi\over a} -
q_{j_{l}}])}\Bigr)\Bigr]\nonumber
\\
& \times &
\Bigl[\prod_{\nu''=1}^{\infty}\,\Bigl(\sum_{j_{1}<j_2<...<j_{N_{s\nu''}}}\sum_{{\cal{P}}}\,(-1)^{\cal{P}}
\,e^{\textstyle
(ia_{s\nu}\sum_{j'=1}^{N_{s\nu''}}j_{{\cal{P}}(l)}\,q_{j_{l}})}\Bigl)\Bigr]\nonumber
\\
& \times &
\vert(L_{c,-1/2},\,L_{s,-1/2});\,(x_{j_{1}},\,x_{j_2},...,x_{j_{N_c}});\,\{(x_{j_{1}},\,x_{j_2},...,
x_{j_{N_{c\nu}}})\} ;\,\{(x_{j_{1}},\,x_{j_2},...,x_{j_{N_{s\nu}}})\}\rangle \, .
\label{wffin}
\end{eqnarray}
The permutations ${\cal{P}}$ on the right-hand side of Eq. (\ref{wffin}) generate a
Slater determinant of the bare momenta and the spatial coordinates of the $N_c$ $c$
pseudoparticles and $N_{\alpha\nu}$ $\alpha\nu$ pseudoparticles belonging to $\alpha\nu$
branches with finite occupancy in the corresponding local charge and spin sequences. On
the left-hand side of Eq. (\ref{wffin}) the set of pseudoparticle occupied bare-momentum
values specify the energy eigenstate. There, $(q_{j_1},\,q_{j_2},...,q_{j_{N_c}})$ are
the set of $N_c$ occupied bare momentum values out of the $N_a$ available discrete $q_j$
values of the $c$ pseudoparticle band such that $j=1,...,N_a$. Moreover,

\begin{equation}
\{(q_{j_1},\,q_{j_2},...,q_{j_{N_{\alpha\nu}}})\} =
(q_{j_1},\,q_{j_2},...,q_{j_{N_{\alpha 1}}});\, (q_{j_1},\,q_{j_2},...,q_{j_{N_{\alpha
2}}});\,(q_{j_1},\,q_{j_2},..., q_{j_{N_{\alpha 3}}});... \, , \label{eanqj}
\end{equation}
are the set of $N_{\alpha\nu}$ occupied bare momentum values out of the $N^*_{\alpha\nu}$
available discrete $q_j$ values of the $\alpha\nu$ pseudoparticle bands of $\alpha\nu$
branches with finite occupancy. Such bare-momentum pseudoparticle occupancy
configurations are discussed in Ref. \cite{I}.

Since equations (\ref{prima}) and (\ref{ellss}) refer to two different representations of
the same states, the energy eigenstates (\ref{wffin}) are Fourier-transform
superpositions involving permutations of the local charge-sequence and spin-sequence
states (\ref{ellss}). The latter local states correspond to rotated-electron site
distribution configurations. Thus, combination of the general expressions
(\ref{completes}) and (\ref{wffin}) provides important information about the relationship
between the bare-momentum pseudoparticles and the rotated-electron occupancy
configurations. We emphasize that the permutations on the right-hand side of Eq.
(\ref{wffin}) are associated with different positions for the local $\alpha\nu$
pseudoparticles but that the internal structure of these quantum objects remains
invariant under these permutations.  Both the Fourier transforms and permutations of the
general expression (\ref{wffin}) do not touch the internal structure of the local
$\alpha\nu$ pseudoparticles and refer only to their translational degrees of freedom.
This results from the point-like character of the occupied sites of the effective
pseudoparticle lattices. Thus, the local and bare-momentum $\alpha\nu$ pseudoparticles
have the same internal structure.

Also the rotated electrons have alternative local and momentum representations. The
creation operator for a rotated electron of momentum $k$ and spin $\sigma$ is such that
${\tilde{c}}_{k,\,\sigma}^{\dag} =
{\hat{V}}^{\dag}(U/t)\,c_{k,\,\sigma}^{\dag}\,{\hat{V}}(U/t)$. Here
$c_{k,\,\sigma}^{\dag}$ refers to the corresponding electronic operator and
${\hat{V}}(U/t)$ is the electron - rotated-electron unitary operator ${\hat{V}}(U/t)$
uniquely defined by Eqs. (21)-(23) of Ref. \cite{I}. The rotated-electron site
distribution configurations studied in previous sections of this paper refer to local
rotated electrons. As the local rotated electrons and local pseudoparticles are directly
related, one could also express the bare-momentum pseudoparticles in terms of momentum
rotated electrons. However, for the energy eigenstates given in Eq. (\ref{wffin}) the
simplest way to relate the pseudoparticles to the rotated electrons is through the local
representations of these quantum objects. In this paper we clarified how the local
pseudoparticles are related to local rotated electrons. Since the bare-momentum
pseudoparticle description is related to the corresponding local pseudoparticle
representation by a simple Fourier transform, such a transform also relates the
bare-momentum pseudoparticles to the local rotated electrons.

For the energy eigenstates given in Eq. (\ref{wffin}), a first condition for the
fulfilment of the periodic boundary conditions of the original electronic problem is that
in all the local charge and spin sequences described by the states (\ref{completes}), the
local $\alpha\nu$ pseudoparticles are described by the same rotated-electron $2\nu$-site
distribution configurations given in Eqs. (\ref{cnp}) and (\ref{snp}) for $\alpha =c$ and
$\alpha =s$, respectively. These are obtained from the proper symmetrization of the
corresponding single pseudoparticle $2\nu$-site sequences. A second condition to ensure
these periodic boundary conditions is that the energy eigenstates given in Eq.
(\ref{wffin}) are Fourier transform superpositions of these charge and spin sequences of
the form given on the right-hand side of such an equation. Provided that the discrete
values of the bare momentum $q_j$ of these Slater determinant superpositions obey Eqs.
(\ref{pbccp}) and (\ref{pbcanp}), the periodic boundary conditions of the original
electronic problem are ensured. The concept of an effective pseudoparticle lattice
introduced above allows the following physical interpretation of the pseudoparticle
bare-momentum boundary conditions (\ref{pbccp}) and (\ref{pbcanp}) needed to ensure the
periodic boundary conditions of the original electronic problem. When the number
$\sum_{\alpha=c,\,s}\sum_{\nu=1}^{\infty}N_{\alpha\nu}$ is odd (and the number of sites
$N^*_{\alpha\nu}$ is even), Eq. (\ref{pbccp}) (and (\ref{pbcanp})) just imposes a twisted
boundary condition such that each $c$ pseudoparticle (and $\alpha\nu$ pseudoparticle)
hopping from site $N_a$ to site $0$ (and site $N^*_{\alpha\nu}$ to site $0$) of the
corresponding effective lattice will acquire a phase factor given by $e^{i\pi}=-1$ or
$e^{i0}=1$. Such a phase factor corresponds to the scattering-less part of the $c$
pseudofermion (and $\alpha\nu$ pseudofermion) $S$ matrix introduced in Ref. \cite{S},
which controls the unusual finite-energy spectral properties of the model \cite{V}.

Note that the general expression provided in Eq. (\ref{wffin}) includes suitable
permutations of the local sequences. In the same equation the rotated-electron site
distribution configurations of the local charge and spin sequences associated with the
states (\ref{prima}) or (\ref{completes}) are reexpressed in terms of pseudoparticle
occupancy configurations of the corresponding effective pseudoparticle lattices. Finally,
let us consider the specific case of a ground state with electronic densities and spin
densities in the ranges $0\leq na\leq 1$ and $0\leq ma\leq na$, respectively. Such a
state is a superposition of local normalized states (\ref{completes}) which have no
$-1/2$ Yang holons, $-1/2$ HL spinons, $c\nu$ pseudoparticles, and $s\nu$ pseudoparticles
belonging to branches such that $\nu>1$ \cite{I,II,V}. Thus, for such a ground state
these local states are of the following simplified form,

\begin{eqnarray}
& & 2^{N_{s1}/2}\,
\vert(0,\,0);\,(x_{j_{1}},\,x_{j_2},...,x_{j_{N_c}});\,(x_{j_{1}},\,x_{j_2},...,
x_{j_{N_{s1}}})\rangle
\nonumber \\
& = & \Bigl[\prod_{j'=1}^{N_{s1}}
(1-{\hat{\cal{T}}}_{s1,\,j',\,1})\Bigr]\,\vert(0,\,0);\,(j_{1},\,j_2,...,j_{N_c});
\,\{(\downarrow_{l_{j',\,1}},\,\uparrow_{l_{j',\,2}})\}\rangle \, , \label{localGS}
\end{eqnarray}
where

\begin{equation}
\{(\downarrow_{l_{j,\,1}},\,\uparrow_{l_{j,\,2}})\} =
(\downarrow_{l_{1,\,1}},\,\uparrow_{l_{1,\,2}}),\,(\downarrow_{l_{2,\,1}},\,\uparrow_{l_{2,\,2}}),...,
(\downarrow_{l_{N_{s1}},\,1},\,\uparrow_{l_{N_{s1},\,2}}) \, , \label{argGSl}
\end{equation}
and the operators ${\hat{\cal{T}}}_{s1,\,j'',\,1}$ are the same as in the general Eq.
(\ref{completes}). The states defined in Eqs. (\ref{localGS}) and (\ref{argGSl}) provide
the position of the two rotated-electron singly occupied sites associated with the
internal structure of each local $s1$ pseudoparticle. These local states have an
alternative representation in terms of the $c$ and $s1$ pseudoparticle occupancy
configurations of the effective $c$ and $s1$ pseudoparticle lattices, respectively.
Within such an effective pseudoparticle lattice representation the local states
(\ref{localGS}) are denoted by $\vert(0,\,0);\,(x_{j_{1}},\,x_{j_2},...,x_{j_{N_c}})
;\,(x_{j_{1}},\,x_{j_2},...,x_{j_{N_{s1}}})\rangle$. Here
$(x_{j_{1}},\,x_{j_2},...,x_{j_{N_{s1}}})$ are the spatial coordinates which define the
positions of the local $s1$ pseudoparticles in the effective $s1$ pseudoparticle lattice.
These spatial coordinates correspond to the indices $(l_{1},\,l_2,...,l_{N_{s1}})$ on the
left-hand side of Eq. (\ref{localGS}). The sites of such an effective $s1$ pseudoparticle
lattice are equally spaced, the corresponding ground-state lattice constant $a_{s1}^0$
being given in Eq. (\ref{acanGS}). The ground state expression is a particular case of
the Fourier-transform superpositions of local states given in Eq. (\ref{wffin}). For the
ground state such an expression simplifies to,

\begin{eqnarray}
\vert GS\rangle & = &
N_a^{N^0/2}\,(N^0_{\uparrow})^{N^0_{\downarrow}/2}\vert(0,\,0);\,(q_{j_1},\,q_{j_2},...,q_{j_{N^0}})
;\,(q_{j_1},\,q_{j_2},...,q_{j_{N^0_{\downarrow}}})\rangle
\nonumber \\
& = & \sum_{c_{1}<c_2<...<c_{N^0}}\sum_{{\cal{P}}}\,(-1)^{\cal{P}} \,e^{\textstyle
(ia\sum_{l=1}^{N^0}c_{{\cal{P}}(l)}\,q_{j_{l}})}\,
\sum_{g_{1}<g_2<...<g_{N^0_{\downarrow}}}\sum_{{\cal{P}}}\,(-1)^{\cal{P}} \,e^{\textstyle
(ia_{s1}\sum_{j'=1}^{N^0_{\downarrow}}g_{{\cal{P}}(l)}\,q_{j_{l}})}\nonumber
\\
& \times & \vert(0,\,0);\,(x_{j_{1}},\,x_{j_2},...,x_{j_{N^0}})
;\,(x_{j_{1}},\,x_{j_2},...,x_{j_{N^0_{\downarrow}}})\rangle \, , \label{GSfin}
\end{eqnarray}
where $N_c =N$, $N_{s1}=N_{\downarrow}$, and $N^*_{s1}=N_{\uparrow}$.

\section{DISCUSSION AND CONCLUDING REMARKS}

The electronic site distribution configurations which describe the energy eigenstates of
the 1D Hubbard model are very complex and dependent on the value of $U/t$. This is
confirmed by the $U/t$ dependence of the double-occupation quantities studied in Ref.
\cite{II}. However, the electron - rotated-electron unitary transformation is such that
it maps these complex and $U/t$ dependent electronic site distribution configurations
onto the corresponding $U/t$ independent rotated-electron site distribution
configurations studied in this paper. We found here that the local pseudoparticle site
distribution configurations of the effective pseudoparticle lattices which describe the
same states are also independent of the value of $U/t$. Moreover, since the spatial
coordinate of these effective lattices is the conjugate of the pseudoparticle
bare-momentum, the pseudoparticle bare-momentum occupancies which describe the energy
eigenstates are also independent of the values of $U/t$. As discussed in the previous
section, this independence on the value of $U/t$ is related to the invariance of the
electronic lattice and the momentum operator (\ref{Popel}), which is the generator of the
spatial translations in such a lattice, under the electron - rotated-electron unitary
transformation.

In this paper we profitted from such a $U/t$ independence of the pseudoparticle and
rotated-electron occupancy configurations and extracted some of our results from analysis
of the corresponding $U/t\rightarrow\infty$ electronic problem. In such a limit the
rotated electron and electron are the same object and the description of the quantum
problem simplifies because both the $\eta$-spin and spin occupancy configurations are
highly degenerated and electron double occupation is a good quantum number, as discussed
in the Appendix. We presented a detailed study of the relation of the description of the
energy eigenstates of the model in terms of rotated-electron site distribution
configurations to the representation of the same states in terms of Yang holon, HL
spinon, and pseudoparticle bare-momentum occupancy configurations. The latter
representation of the energy eigenstates is obtained from the Bethe-ansatz solution and
$\eta$-spin and spin algebras. The connection between these two descriptions follows from
the relation of rotated electrons to the quantum numbers of the above solution and
symmetries established in Ref. \cite{I}. That connection involves complex
rotated-electron behavior such that, for {\it all} energy eigenstates, the $N_c$ rotated
electrons of the singly occupied sites separate into $M_s=N_c$ charge-less spin $1/2$
spinons and $N_c$ spin-less and $\eta$-spin-less $c$ pseudoparticles of charge $-e$. (The
$-1/2$ and $+1/2$ holons are less exotic and correspond to the rotated-electron
doubly-occupied and unoccupied sites, respectively.) To further study such a connection,
in this paper we introduced the concepts of a local pseudoparticle and an effective
pseudoparticle lattice. For an effective $\alpha\nu$ pseudoparticle lattice the values of
its length and lattice constant have an exotic dependence on the values of the numbers of
quantum objects in a given state but are independent of the value of $U/t$. The
corresponding pseudoparticle spatial coordinate $x_j$ is given in Eqs. (\ref{xc}) and
(\ref{xeg}) and is the conjugate of the pseudoparticle bare-momentum $q_j$. The
pseudoparticle bare-momentum is a quantum number which emerges from the Bethe-ansatz
Takahashi's thermodynamic equations \cite{Takahashi,I}. The introduction of the concept
of an effective pseudoparticle lattice involved the study of the rotated-electron
distribution configurations of doubly occupied and unoccupied sites (and spin-down and
spin-up singly occupied sites) which describe the internal structure of a local $c\nu$
pseudoparticle (and $s\nu$ pseudoparticle).

In what the translational degrees of freedom are concerned, the local $\alpha\nu$
pseudoparticle is a point-like quantum object, its internal structure being the same as
that of the corresponding bare-momentum pseudoparticle. We also found that the spatial
coordinates of the occupied and unoccupied sites of the effective $c$ pseudoparticle
lattice are the same as the corresponding coordinates of the sites singly occupied by
rotated electrons and doubly occupied by or free of rotated electrons, respectively. The
energy eigenstates can be described in terms of local pseudoparticle site distribution
configurations in the corresponding effective pseudoparticle lattice. Our results reveal
that there is an one-to-one correspondence between the local pseudoparticle site
distribution configurations in the effective pseudoparticle lattices which describe a
given energy eigenstate and the rotated-electron site distribution configurations which
describe the same state.

The results of this paper provide useful information for the correct identification of
the scatterers and scattering centers \cite{S} that control the model unusual spectral
properties \cite{V}. The use of the pseudoparticle description studied in this paper for
the evaluation of the finite-energy spectral-weight distributions \cite{V} involves a
second unitary transformation, which maps the $c$ pseudoparticles (and composite $c\nu$
or $s\nu$ pseudoparticles) onto $c$ pseudofermions (and composite $c\nu$ or $s\nu$
pseudofermions) and is defined in the Hilbert subspace where the one- and two-electron
excitations are contained \cite{IIIb}. That transformation introduces shifts of order
$1/L$ in the pseudoparticle discrete momentum values and leaves all other pseudoparticle
properties invariant. (The Yang holons and HL spinons remain invariant under such a
transformation.) Since the effective pseudoparticle lattices and the corresponding site
distribution configurations that describe the energy eigenstates belonging to that
subspace remain invariant under the pseudoparticle - pseudofermion unitary
transformation, our results concerning the local site distribution configurations apply
both to pseudoparticles and pseudofermions.

As for the pseudoparticles, the unoccupied sites of the effective pseudofermion lattices
correspond to the sites left over by the occupied sites and vice versa. Thus, we can
uniquely specify a given effective pseudfermion lattice site distribution configuration
by providing the position of the occupied sites (or unoccupied sites) only. Moreover, it
follows from our detailed studies that although the number of occupied (and unoccupied)
sites of the $c$ pseudofermions equals the number of spinons, $N_c=M_s$ (and the number
of holons, $N^h_c =M_c$) a $c$ pseudofermion (and $c$ pseudofermion hole) is not a spinon
(and a holon). Our results also reveal that the fact that the number of $s\nu$
pseudofermion holes (and $c\nu$ pseudofermion holes) is given by $N^h_{s\nu} = L_{s} +
2\sum_{\nu'=\nu +1}^{\infty} (\nu' -\nu) N_{s\nu'}$ (and $N^h_{c\nu} = L_{c} +
2\sum_{\nu'=\nu +1}^{\infty} (\nu' -\nu) N_{c\nu'}$) does not imply that the $N^h_{s\nu}$
$s\nu$ pseudofermion holes (and $N^h_{c\nu}$ $c\nu$ pseudofermion holes) of a given
energy eigenstate are the same objects as the $L_{s}$ HL spinons and $(\nu' -\nu)$ $-1/2$
spinons and $(\nu' -\nu)$ $+1/2$ spinons of the set of composite $s\nu'$ pseudofermions
of the same state such that $\nu'>\nu$ (and the $L_{c}$ Yang holons and $(\nu' -\nu)$
$-1/2$ holons and $(\nu' -\nu)$ $+1/2$ holons of the set of composite $c\nu'$
pseudofermions of the same state such that $\nu'>\nu$). Indeed, both the HL spinon and
Yang holon occupancies and the $\alpha\nu$ pseudofermion spatial coordinate occupancy
configuration of each $\alpha\nu$ pseudofermion branch with finite occupancy in a given
energy eigenstate are independently conserved.

Since the Yang holons and HL spinons remain invariant under the pseudoparticle -
pseudofermion unitary transformation they are neither scatterers nor scattering centers,
whereas both the pseudofermion and pseudofermion holes are scatterers, and the
pseudofermions and pseudofermion holes created in a ground-state - excited-state
transition are the scattering centers \cite{S}. The results of this paper confirm that
the $c$ pseudofermion holes are spin-less and $\eta$-spin-less objects, as the
corresponding $c$ pseudofermions, in spite of their number equaling the number of
$\eta$-spin $1/2$ holons. This is consistent with the complementary studies of Ref.
\cite{I}, which reveal that the $c$ pseudofermion and hole occupancy configurations are
not related to the $\eta$-spin irreducible representations. Our results also confirm that
the $s1$ pseudofermion holes have zero spin, as the corresponding $s1$ pseudofermions, in
spite of their number reading $N^h_{s1} = L_{s} + 2\sum_{\nu'=2}^{\infty} (\nu' -1)
N_{s\nu'}$. This is again consistent with the complementary studies of Ref. \cite{I},
which show that the $s1$ pseudofermion and hole occupancy configurations only contribute
to the spin singlet occupancy configurations. That the $c$ pseudofermions and $c$
pseudofermion holes are $\eta$-spin-less and spin-less objects and the $s1$
pseudofermions and $s1$ pseudofermion holes carry zero spin justifies the simple form of
the corresponding $S$ matrices, which have dimension one \cite{S}. Such $S$ matrices
fully control the unusual spectral properties of the model through the pseudofermion
anticommutators \cite{V,S}. If they had dimension larger than one, the pseudofermion
anticommutators would also be matrices of dimension larger than one and the study of the
spectral properties would be much more involved. Furthermore, the concepts of a local
pseudoparticle (and pseudofermion) and an effective pseudoparticle (and pseudofermion)
lattice introduced here are applied in the studies of Ref. \cite{V} to the evaluation of
finite-energy spectral function expressions. Therefore, the studies of this paper provide
new insights and contribute to the further understanding of the unusual finite-energy
properties observed in low-dimensional materials \cite{super,spectral0,V,spectral}.

\begin{acknowledgments}
I thank Daniel Bozi, Patrick A. Lee, Gerardo Ortiz, Karlo Penc, and Pedro Sacramento for
stimulating discussions and Nuno Peres for useful discussions related to the occupancy
configurations represented in the figures of this paper. I also thank the hospitality and
support of MIT, the financial support of the Gulbenkian Foundation, Fulbright Commission,
and FCT grant POCTI/FIS/58133/2004, and the hospitality and support of the Aspen Center
for Physics, where the last stage of this research was performed.
\end{acknowledgments}
\appendix

\section{PROPERTIES OF THE LARGE $U/t$ PHYSICS USEFUL FOR OUR STUDIES}

In this Appendix we review some aspects of the large $U/t$ physics that are useful for
the problems studied in this paper. By use of Eqs. (12)-16) of Ref. \cite{I}, we find
that in the limit $U/t\rightarrow\infty$ the momentum (\ref{Popel}) and the energy $E_H$
associated with the simple 1D Hubbard Hamiltonian ${\hat{H}}_H $ given in Eq. (\ref{H})
simplify to Eq. (36) of Ref. \cite{I} with $M_{c,\,-1/2}=D$ and expression
(\ref{EHUinf}), respectively. (Here the electronic double occupation $D$ is a good
quantum number such that $D=M_{c,\,-1/2}$ and for finite values of $U/t$ the number
$M_{c,\,-1/2}$ refers to the rotated-electron double occupation.) For zero spin density,
electronic densities in the range $0\leq na\leq 1$, and on-site repulsion $U>>t$ the
ground state energy $E_0$ of the 1D Hubbard model is given by \cite{Carmelo88},

\begin{equation}
E_0 = -{2N_a t\over\pi}\sin (\pi n\,a) -U\,D_0 \, ; \hspace{0.5cm} D_0 = {\partial
E_0\over\partial U} =\Bigl({t\over U}\Bigr)^2\,4N\, n\,\ln 2\,\Bigl(1-{\sin (2\pi
n\,a)\over 2\pi n\,a}\Bigr) \, , \label{E0}
\end{equation}
where $D_0$ is the ground-state expectation value of the electronic double occupation
\cite{II}. Here the term $-[2N_a t/\pi]\sin (\pi n\,a)$ is the kinetic energy associated
with the hopping processes which do not change electronic double occupation. On the other
hand, the energy term $-U\,D_0$ includes both kinetic and potential energy contributions,
$-U\,2D_0$ and $U\,D_0$, respectively. It arises from hopping processes which change
electronic double occupation and lead to the ground-state expectation value $D_0$ given
in Eq. (\ref{E0}). Thus, the physics associated with this second energy term corresponds
to excitation processes of higher order in $t/U$ relative to the $t/U\rightarrow 0$ limit
where electronic double occupation is a good quantum number and the ground-state
electronic double occupation of the model is exactly zero. If we include energy
contributions of the order $t(t/U)^1$, the ground state contains a small but finite
electronic double occupation expectation value given in Eq. (\ref{E0}) and the
corresponding quantum problem is not equivalent to the physical situation of interest for
the rotated-electron studies of this paper. Indeed, electrons equal rotated electrons
when the limit $t/U\rightarrow 0$ is reached and double occupation is a good quantum
number. Only in such a limit do the electronic occupancy configurations which describe
the bare-momentum energy eigenstates equal the corresponding rotated-electron
configurations that are valid for all values of $U/t$.

Note that at half filling the electronic density reads $na=1$, the energy term of order
$t(t/U)^0$ on the right-hand side of Eq. (\ref{E0}) vanishes, and the ground-state energy
and electronic double occupation expressions given in Eq. (\ref{E0}) simplify to
$E_0=-[t^2/U]\,4N_a\, \ln 2$ and $D_0 = (t/U)^2\,4N_a\, \ln 2$, respectively. It is well
known that this energy can be associated with an isotropic Heisenberg model \cite{Emery}.
The corresponding ground state leads to energy contributions of the order $t(t/U)^1$ and
thus contains a small but finite electronic double occupation expectation value, $D_0 =
(t/U)^2\,4N_a\, \ln 2$. It follows that the usual description of the large-$U/t$
half-filling Hubbard model in terms of an isotropic Heisenberg model is not equivalent to
our limit. In such a limit only hopping processes which do not change the value of
electronic double occupation must be considered.

To leading order in the parameter $t/U$, the energy spectrum of the 1D Hubbard model in
the limit $t/U\rightarrow 0$ is of the form given in Eq. (\ref{EHUinf}). The permitted
hopping processes lead to contributions in the eigenstate energies of order $t(t/U)^{-1}$
and $t(t/U)^0$. These contributions are associated with the energy terms $[N_a /2\pi]
\int_{q_c^{-}}^{q_c^{+}} dq\, N_c (q) [-2t\cos q]$ and $U\,D$, respectively, on the
right-hand side of Eq. (\ref{EHUinf}). In the particular case of the ground state there
are no contributions of order $t(t/U)^{-1}$ because the electronic double occupation
eigenvalue $D_0$ is zero. In the limit $t/U\rightarrow 0$ only hopping processes such
that the electronic singly occupied sites can move relatively to the electronic
doubly-occupied and unoccupied site distribution configurations without changing these
configurations are permitted. In such a limit there is a huge degeneracy of $\eta$-spin
and spin occupancy configurations. Thus, there are several choices for complete sets of
energy eigenstates with the same energy and momentum spectra given in Eq. (\ref{EHUinf})
and Eq. (36) of Ref. \cite{I} with $M_{c,\,-1/2}$ replaced by electronic double
occupation, respectively. This justifies why in this limit there are many choices for
complete sets of compatible observables. The 1D Hubbard model in the limit of
$U/t\rightarrow\infty$ has been studied in the literature by many authors
\cite{Penc96-97,Harris,Ogata,Geb,Carmelo88,Beni,Ricardo,Eskes}. For the evaluation of
quantities describing the physics of the model in the limit of $U/t\rightarrow\infty$,
the alternative use of different complete sets of states leads to the same final
expressions for correlation functions and other quantities of physical interest.

In this paper we are interested in one of these choices of energy eigenstates only. It
corresponds to the complete set of $4^{N_a}$ energy eigenstates generated from the
corresponding energy eigenstates of the finite-$U/t$ 1D Hubbard model by turning off
adiabatically the parameter $t/U$. Only such a set of states corresponds to the states
obtained from the finite $t/U$ states by the electron - rotated-electron unitary
transformation. These states are common eigenstates of both the 1D Hubbard model as
$U/t\rightarrow\infty$ and of the set of number operators $\{{\hat{L}}_{\alpha
,\,-1/2}\}$, $\{{\hat{N}}_c (q_j)\}$, and $\{{\hat{N}}_{\alpha\nu}(q_j)\}$ with $\alpha
=c,s$ and $\nu=1,2,3,...$ considered in Ref. \cite{I} in the same limit. Together with
the Hamiltonian, these operators constitute a complete set of compatible and commuting
hermitian operators. These operators also commute with the momentum operator $\hat{P}$.
For these energy eigenstates the bare momentum $q_j$ of the $c$ pseudoparticles and
$c\nu$ and $s\nu$ pseudoparticles such that $\nu=1,2,...$ is a good quantum number. This
justifies the designation of bare-momentum energy eigenstates. However, in the limit
$t/U\rightarrow 0$ there are other choices for complete sets of energy and momentum
eigenstates which are due to the periodic boundary conditions of the original electronic
problem. In section III we discuss an alternative set of energy eigenstates used in the
studies of Ref. \cite{Geb}.

Finally, we briefly discuss the relation of the effective pseudoparticle lattices
introduced in Sec. V to previous results on the model in the limit
$U/t\rightarrow\infty$. It is well known that in such a limit and for the subspace with
no electronic double occupancy and thus such that $N_c=N$ and at zero spin density the
charge and spin degrees of freedom of the model can be described by two independent
systems of $N$ spin-less fermions and $N/2$ spin-down spins \cite{Ogata,Penc96-97,I,II}.
The spin-less fermions can be associated with an effective lattice with $j=1,2,3,...,N_a$
sites, whereas the $N/2$ spin-down spins correspond to a squeezed effective lattice with
$j=1,2,3,...,N/2$ sites. The spin-less fermion and spin occupancy configurations of these
effective lattices describe electronic site distribution configurations. In the limit
$U/t\rightarrow\infty$ such spin-less fermion and spin effective lattices are directly
related to the effective $c$ pseudoparticle and $s1$ pseudoparticle lattices,
respectively, introduced in Sec. V. In the limit $U/t\rightarrow\infty$ the energy
eigenstates with finite occupancies in the effective $c\nu$ pseudoparticle lattices have
an infinite energy relative to the ground state and do not contribute to the
finite-energy physics. This explains why these states are unimportant for the
$U/t\rightarrow\infty$ physics and are in general not considered \cite{Ogata,Penc96-97}.
In the limit $U/t\rightarrow\infty$ the spin excitations involving both occupancy
configurations of $s1$ pseudoparticles and $s\nu$ pseudoparticles belonging to branches
such that $\nu>1$ are often described by the isotropic Heisenberg chain which describes
these excitations \cite{Ogata,Penc96-97}.


\end{document}